\documentclass{aa} 

\usepackage{graphicx}
\usepackage[english]{babel}
\usepackage{natbib}
\usepackage[pdftex]{hyperref}
\usepackage{amsmath}
\usepackage{txfonts}

\bibpunct{(}{)}{;}{a}{}{,}

\begin{document}

\title{Population synthesis of ultracompact X-ray binaries\\ in the Galactic Bulge}

\author{
L.~M.~van~Haaften\inst{\ref{radboud}} \and
G.~Nelemans \inst{\ref{radboud},\ref{leuven}} \and
R.~Voss\inst{\ref{radboud}} \and
S.~Toonen\inst{\ref{radboud}} \and
S.~F.~Portegies~Zwart\inst{\ref{leiden}} \and
L.~R.~Yungelson\inst{\ref{inasan}} \and
M.~V.~van~der~Sluys\inst{\ref{radboud}}
}

\institute{
Department of Astrophysics/ IMAPP, Radboud University Nijmegen, P.O. Box 9010, 6500 GL Nijmegen, The Netherlands, \email{L.vanHaaften@astro.ru.nl} \label{radboud} \and
Institute for Astronomy, KU Leuven, Celestijnenlaan 200D, 3001 Leuven, Belgium \label{leuven} \and
Leiden Observatory, Leiden University, P.O. Box 9513, 2300 RA Leiden, The Netherlands \label{leiden} \and
Institute of Astronomy, Russian Academy of Sciences, 48 Pyatnitskaya Str., 119017 Moscow, Russia \label{inasan}
}

\abstract {} 
{We model the present-day number and properties of ultracompact X-ray binaries (UCXBs) in the Galactic Bulge. The main objective is to compare the results to the known UCXB population as well as to data from the Galactic Bulge Survey, in order to learn about the formation of UCXBs and their evolution, such as the onset of mass transfer and late-time behavior.} 
{The binary population synthesis code \textsf{SeBa} and detailed stellar evolutionary tracks have been used to model the UCXB population in the Bulge. The luminosity behavior of UCXBs has been predicted using long-term X-ray observations of the known UCXBs as well as the thermal-viscous disk instability model.} 
{In our model, the majority of UCXBs initially have a helium burning star donor. Of the white dwarf donors, most have helium composition. In the absence of a mechanism that destroys old UCXBs, we predict $(0.2 - 1.9) \times 10^{5}$ UCXBs in the Galactic Bulge, depending on assumptions, mostly at orbital periods longer than $60$ min (a large number of long-period systems also follows from the observed short-period UCXB population). About $5 - 50$ UCXBs should be brighter than $10^{35}\ \mbox{erg s}^{-1}$, mostly persistent sources with orbital periods shorter than about $30$ min and with degenerate helium and carbon-oxygen donors. This is about one order of magnitude more than the observed number of (probably) three.} 
{This overprediction of short-period UCXBs by roughly one order of magnitude implies that fewer systems are formed, or that a super-Eddington mass transfer rate is more difficult to survive than we assumed. The very small number of observed long-period UCXBs with respect to short-period UCXBs, the surprisingly high luminosity of the observed UCXBs with orbital periods around $50$ min, and the properties of the PSR J1719--1438 system all point to much faster UCXB evolution than expected from angular momentum loss via gravitational wave radiation alone. Old UCXBs, if they still exist, probably have orbital periods longer than $2$ h and have become very faint due to either reduced accretion or quiescence, or have become detached. UCXBs are promising candidate progenitors of isolated millisecond radio pulsars.} 

\authorrunning{L.~M.~van~Haaften et al.}
\titlerunning{Population synthesis of UCXBs in the Galactic Bulge}

\maketitle

\section{Introduction}
\label{sect:ucxb:intro}

Ultracompact X-ray binaries (UCXBs) are low-mass X-ray binaries with observed orbital periods shorter than $\sim \! 1$ h, indicating a compact, hydrogen deficient donor star \citep{vila1971,paczynski1981cv,sienkiewicz1984}. The donor overflows its Roche lobe, and lost matter is partially accreted by a neutron star or black hole companion. Because of the compact orbit, mass transfer is driven by orbital angular momentum loss via gravitational wave radiation \citep[e.g.][]{kraft1962,paczynski1967b,faulkner1971,pringle1975,tutukov1979}. For an overview of the relevance of studying UCXBs, see e.g. \citet{nelemans2010b}.

The Galactic Bulge is a suitable environment to discover UCXBs because of the high local star concentration. Furthermore, the Bulge is an old stellar population that contains few young X-ray sources such as high-mass X-ray binaries and core-collapse supernova remnants. Due to their rarity, UCXBs are not found nearby in the Galaxy. Also, while observable in X-rays, they are too faint to be identified in other galaxies \citep[e.g.][]{voss2007m31}.

The Galactic Bulge Survey (GBS) \citep{jonker2011} is an X-ray and optical survey focused on two $6^{\circ} \times 1^{\circ}$ regions centered $1.5^{\circ}$ to the North and South of the Galactic Center. One of the goals of the GBS is to investigate the properties of populations of X-ray binaries in order to constrain their formation scenarios, especially the common-envelope phase(s).

The present study aims to predict and explain GBS results regarding the number and luminosities of UCXBs by means of binary population synthesis and UCXB evolutionary tracks, thereby contributing to a better understanding of the formation and evolution of UCXBs. We also compare our results to the orbital periods and chemical compositions of the known UCXB population. In Sect.~\ref{sect:ucxb:method} we describe our assumptions on the star formation, stellar and binary evolution, and the observable characteristics of evolved UCXBs. The results follow in Sect.~\ref{sect:ucxb:results}, where we present the modeled present-day population and its observational properties. In Sect.~\ref{sect:ucxb:discussion}, we compare our results with the population synthesis studies by \citet{belczynski2004popsynt}, \citet{zhu2012}, and \citet{zhu2012ucxb}, as well as to observations, and we discuss various implications. We conclude in Sect.~\ref{sect:ucxb:conclusion}.

\section{Method}
\label{sect:ucxb:method}

The study of the evolution of the UCXB population consists of several steps.
First, the star formation history of the Galactic Bulge and the binary initial mass function prescribe which types of zero-age main sequence binaries form in the Bulge, and when they form.
The binary population synthesis code \textsf{SeBa} \citep{portegieszwart1996,nelemans2001a,toonen2012} is used to simulate the evolution of this population of zero-age main sequence binaries. During the evolution of the population, all UCXB progenitors are selected at a certain moment after the supernova explosion that leaves behind the neutron star or black hole. More detailed evolutionary tracks are used to trace the subsequent evolution. This yields the present-day number and intrinsic parameters of the UCXBs in the Galactic Bulge.
Finally, using long-term observations by the \emph{Rossi X-ray Timing Explorer} All-Sky Monitor (\emph{RXTE} ASM) \citep{bradt1993,levine1996} as well as the accretion disk instability model, the modeled UCXB parameters are translated into observational parameters using the results of \citet{vanhaaften2012xlum}. The result is a prediction of the present-day observable population.

\subsection{Population synthesis}
\label{sect:ucxb:seba}

The binary population synthesis code \textsf{SeBa} models the evolutionary transformations of a population of binary stars based on a distribution of initial binary parameters. It follows the evolution of stellar components using analytic formulas by \citet{hurley2000}, taking into account circularization due to tidal interaction, magnetic braking, gravitational wave radiation, mass exchange via Roche-lobe overflow, common envelopes, and empirical parameterizations of wind mass loss. For a more extensive description of \textsf{SeBa} we refer to \citet{portegieszwart1996}, \citet{nelemans2001a}, and \citet{toonen2012}.

As is common, we use parameterizations to describe the common-envelope process. In this study, the density profile of the donor envelope is parametrized by $\lambda = 1/2$, the efficiency with which orbital energy is used on unbinding the common envelope by $\alpha_\mathrm{CE} = 4$ (justified by explosive shell burning in massive stars during the common-envelope phase, \citealp{podsiadlowski2010}) and the specific angular momentum of the envelope after it has left the system, relative to the specific angular momentum of the pre-common-envelope binary, by $\gamma = 7/4$. In choosing a value of $\alpha_\mathrm{CE} \lambda = 2$ for massive stars we follow \citet{portegieszwart1998,yungelson2006,yungelson2008lasota}. The $\gamma$- and $\alpha_\mathrm{CE}$-prescriptions are used as described in \citet{toonen2012}. A metallicity $Z = 0.02$ has been used \citep{zoccali2003}. The metallicity is held constant in the population synthesis simulations as well as the subsequent tracks because of the relatively short episode of star formation (Sect.~\ref{sect:ucxb:sfhini}).
For the kicks acquired by nascent neutron stars we use the velocity distribution suggested by \citet{paczynski1990} with a dispersion of $270\ \mbox{km s}^{-1}$.
Because these parameters are very uncertain, in Sect.~\ref{sect:ucxb:discussion} we will consider the effect of varying the common-envelope parameters and the supernova kick velocity distribution from our standard values.

\subsection{Star formation history and initial binary parameters}
\label{sect:ucxb:sfhini}

\begin{figure}
\resizebox{\hsize}{!}{\includegraphics{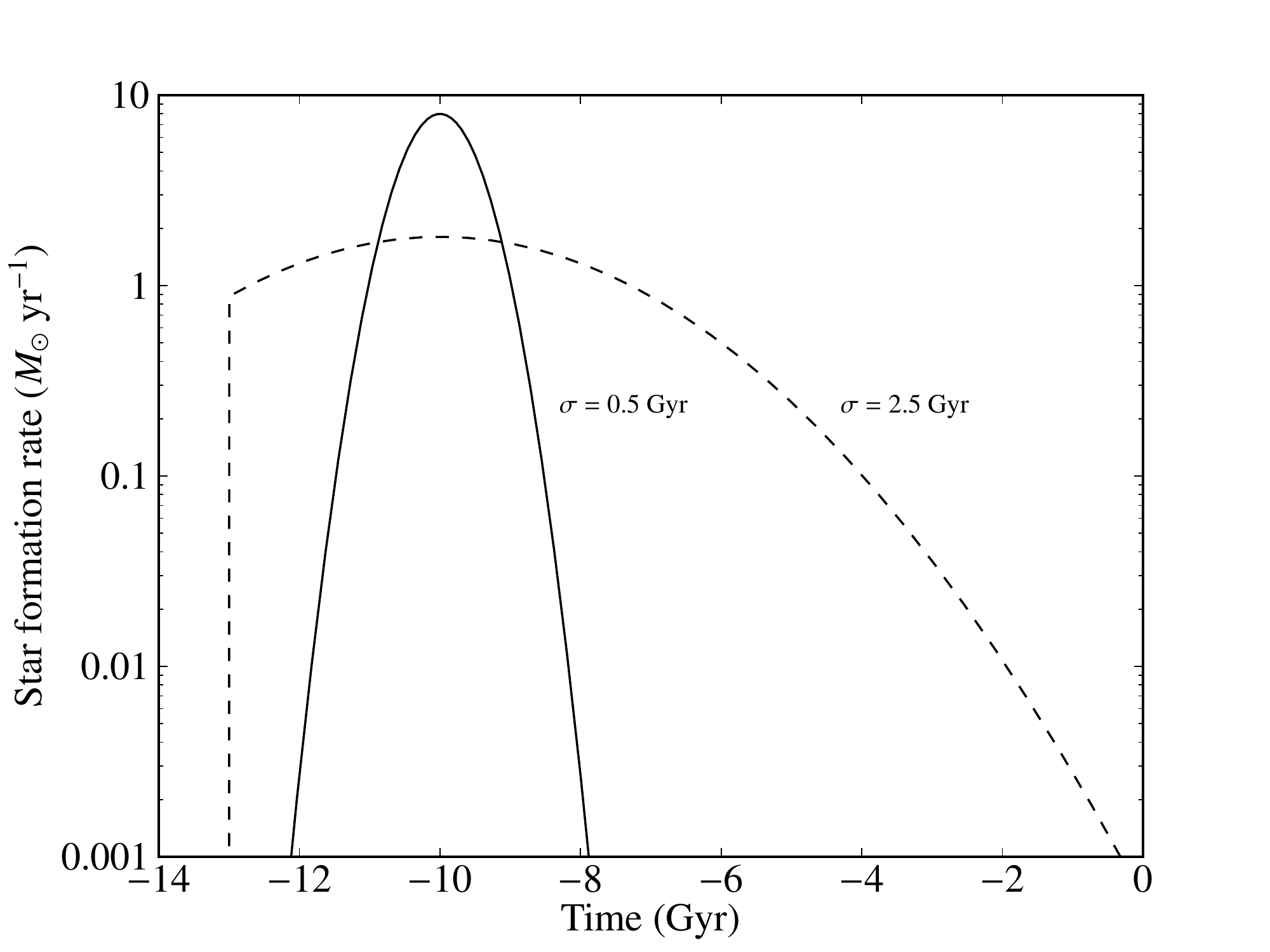}} 
\caption{Star formation history of the Galactic Bulge as a Gaussian distribution for mean $\mu = -10$ Gyr and two values of the standard deviation $\sigma$. The total mass of the stars formed is $1 \times 10^{10}\ M_{\odot}$. Time $= 0$ corresponds to the present. Star formation is assumed to start at Time $= -13$ Gyr.}
\label{fig:ucxb:sfh}
\end{figure}

The star formation history of the Galactic Bulge can be approximated by a Gaussian distribution with a mean $\mu = -10$ Gyr and a standard deviation $\sigma = 0.5 - 2.5$ Gyr, where the total mass of stars that are formed is $1 \times 10^{10}\ M_{\odot}$ \citep{clarkson2009,wyse2009}. Star formation is assumed to start $13$ Gyr before present. For a narrow distribution the star formation is concentrated around $10$ Gyr in the past, but the $\sigma = 2.5$ Gyr distribution has an important tail of recent star formation. In this paper we consider $\sigma = 0.5$ Gyr and $\sigma = 2.5$ Gyr (Fig.~\ref{fig:ucxb:sfh}), representing a relatively instantaneous burst of star formation, and additional recent star formation, respectively.

We find the initial binary parameters by primary-constrained pairing \citep{kouwenhoven2008}. We derive the initial mass function for the primary components by combining the initial mass function for stellar systems by \citet{kroupa2001} with an estimate of the binary fraction as a function of mass. The mass ratio of the secondary and the primary is drawn from a flat distribution between $0$ and $1$. The $1 \times 10^{10}\ M_{\odot}$ of stellar mass is found to contain $5.2 \times 10^{9}$ binary systems. We simulated $1$ million binaries with a lower primary mass limit of $4\ M_{\odot}$ (because systems with a lower primary mass do not produce a supernova event in either component), and another $4.3$ million binaries with a lower primary mass limit of $8\ M_{\odot}$, after it became clear that systems with lower primary masses do not produce UCXBs. Using the binary initial mass function we calculated to how many binaries in the full mass range of primaries ($0.08 - 100\ M_{\odot}$) this simulation corresponds. The resulting population was then multiplied by a factor (of $14.45$) to scale to the entire Bulge population.

For an analytic derivation of the binary initial mass function we refer to Appendix \ref{sect:ucxb:appimf}.

\subsection{Formation scenarios}
\label{sect:ucxb:formation}

We consider three UCXB-progenitor classes, each defined by the stellar type of the donor at the time it fills, or will fill, its Roche lobe:\\
\textbf{Class 1.} White dwarf with a neutron star or black hole companion \citep{tutukov1993,iben1995,yungelson2002},\\
\textbf{Class 2.} Helium burning star with a neutron star or black hole companion \citep{savonije1986,iben1987,yungelson2008},\\
\textbf{Class 3.} Evolved main sequence star of about $1\ M_{\odot}$ with a neutron star or black hole companion \citep{tutukov1985,nelson1986,fedorova1989,pylyser1989,podsiadlowski2002,nelson2003,sluys2005a,lin2011}.

These classes include all the binary systems that may eventually evolve into an UCXB \citep{belczynski2004popsynt,sluys2005a,nelemans2010} -- in some models involving accretion-induced collapse of a white dwarf or a neutron star, the donor star has already transferred mass before the formation of the eventual accretor, a neutron star or a black hole, respectively. The detailed tracks follow the evolution of the helium burning donor starting immediately after its formation, and the main sequence donor immediately after the supernova event. The white dwarf donor tracks start at the onset of Roche-lobe overflow. In each of the detailed tracks the mass transfer is conservative as long as the mass transfer rate does not exceed the Eddington limit. If the mass transfer is faster than that, accretion at the Eddington limit is assumed, and the mass that is lost from the system carries the specific angular momentum of the accretor.

The initial system parameters (component masses and orbital periods) and major events during the evolution towards an UCXB are described below for each class.

\subsubsection{White dwarf donor systems}
\label{sect:ucxb:wdire}

The evolution of UCXBs with a white dwarf donor can be divided into two main categories based on whether the primary (initially more massive star) or secondary component becomes a supernova. A supernova explosion of the secondary star is possible when it gains mass by hydrogen accretion from the primary \citep[e.g.][]{tutukov1993,portegieszwart1996,vankerkwijk1999,portegieszwart1999,tauris2000sen}. A supernova explosion of the secondary probably never produces a black hole, neither does the primary turn into a helium white dwarf after the supernova, because it starts out too massive. Thus, all secondary-supernova systems have a carbon-oxygen or oxygen-neon white dwarf donor and a neutron star accretor. Because the high stellar mass required for a supernova explosion of the primary is relatively rare due to the steep initial mass function, in our simulations a significant fraction of the systems ($13\%$ of the carbon-oxygen white dwarf systems and $36\%$ of the oxygen-neon white dwarf systems) experience their supernova in the secondary star. Systems with a black hole accretor are rare, about $0.2\%$ of all white dwarf systems. All black holes form from the primary and have a $\sim \! 0.6 - 0.8\ M_{\odot}$ carbon-oxygen white dwarf companion.

Carbon-oxygen white dwarf systems\footnote{The envelopes of some white dwarfs contain helium, but we refer to them as carbon-oxygen white dwarfs because the envelope will be lost relatively early.} are $1.5$ times more prevalent than oxygen-neon white dwarf systems in our simulations, and $\sim \! 30$ times more prevalent than helium white dwarf systems (combining primary and secondary supernovae).

\paragraph{Supernova explosion of the primary}

This category can be subdivided by the predominant white dwarf composition: helium, carbon-oxygen or oxygen-neon.

Evolution starts with a zero-age main sequence binary in which the primary is a massive star ($M \gtrsim 8\ M_{\odot}$ if the secondary is to become a helium or carbon-oxygen white dwarf, and $M \gtrsim 10\ M_{\odot}$ in the case of an oxygen-neon white dwarf companion) that evolves off the main sequence first. Systems that eventually produce a helium white dwarf donor have initial orbital periods ranging mainly from $1$ to $100$ yr. For systems that produce a carbon-oxygen or oxygen-neon white dwarf donor the orbital periods lie mostly between $0.1$ and $1000$ yr. In the case of neutron star accretors, the primary expands during the Hertzsprung gap or as a giant and fills its Roche lobe, followed by mass transfer to the companion. The secondary cannot accrete all of this mass and is engulfed in a common envelope \citep{paczynski1976}. The envelope is expelled before the two components merge and the exposed helium core and the main sequence secondary are left behind with an orbital separation several tens of times smaller than before. After a few $10$ Myr, the helium star turns into a giant (which may lead to a subsequent phase of Roche-lobe overflow) and explodes as a core-collapse supernova, leaving behind a neutron star. Black hole progenitors are Wolf-Rayet stars, which lose a large fraction of their mass before evolving into a helium giant, and then collapse to form a black hole.

Secondary stars that eventually become a helium white dwarf donor had a zero-age main sequence mass between $1.4$ and $2.3\ M_{\odot}$, whereas for carbon-oxygen white dwarf donors this range is $2.3 - 7\ M_{\odot}$, where $2.3\ M_{\odot}$ is the maximum mass of single stars that undergo the helium flash. A small fraction started with a higher initial mass.
Progenitors of oxygen-neon white dwarf secondaries have a mass between $7$ and $11\ M_{\odot}$ on the zero-age main sequence and do not become a supernova due to severe mass loss \citep[e.g.][]{gilpons2001}. (There is some overlap with the progenitor mass range of carbon-oxygen white dwarfs -- the end product depends on whether burning stops before of after carbon ignition.) After the supernova explosion has occurred, the secondary evolves off the main sequence. As a subgiant, it initiates a common envelope with the neutron star, shrinking the orbit by another factor of a few tens. The core cools into a helium white dwarf ($\lesssim 0.35\ M_{\odot}$) or, after a helium burning and helium giant stage, a carbon-oxygen white dwarf ($0.35 - 1.1\ M_{\odot}$), or even, after carbon burning, an oxygen-neon white dwarf ($\gtrsim 1.1\ M_{\odot}$, \citealp{gilpons2001}). Orbital angular momentum loss via gravitational wave radiation further shrinks the orbit until the white dwarf eventually overfills its Roche lobe, which happens at an orbital period of a few minutes.

\paragraph{Supernova explosion of the secondary}

In this scenario, the total binary mass needs to be at least $9\ M_{\odot}$ if the primary becomes a carbon-oxygen white dwarf and $12\ M_{\odot}$ if the primary becomes an oxygen-neon white dwarf. The primary transfers several solar masses to the secondary in a stable manner (avoiding a common envelope -- initially the secondary must have a mass of at least $0.55$ times the primary mass) while ascending the red giant branch \citep[e.g.][]{tauris2000sen}. Eventually the core becomes a helium star, or a carbon-oxygen or oxygen-neon white dwarf. The secondary, which becomes the more massive component of the system, evolves off the main sequence and initiates a common envelope. The orbit shrinks, and $30 - 70$ Myr after the binary formation the secondary explodes as a supernova and produces a neutron star. In systems which remain bound after the supernova explosion, the primary will eventually reach Roche-lobe overflow as a relatively massive ($\gtrsim 0.7\ M_{\odot}$) carbon-oxygen white dwarf or an oxygen-neon white dwarf ($\gtrsim 1.1\ M_{\odot}$). The relatively high initial mass of the primary precludes less massive carbon-oxygen white dwarfs. The initial orbital period in this scenario can be much shorter than in the primary-supernova scenarios, down to a few days. The initial stellar masses lie between $4.5$ and $\sim \! 10\ M_{\odot}$ for the primary and $4 - 9\ M_{\odot}$ for the secondary (if the former becomes a carbon-oxygen white dwarf) and about $5 - 12\ M_{\odot}$ for both the primary and secondary (if the former becomes an oxygen-neon white dwarf).

\paragraph{The onset of mass transfer from the white dwarf}

Most white dwarf--neutron star systems merge upon the onset of mass transfer. For a $1.4\ M_{\odot}$ neutron star companion, white dwarfs with a mass higher than $\sim \! 0.83\ M_{\odot}$ experience dynamically unstable mass transfer, assuming a zero-temperature (completely degenerate) mass-radius relation for the donor \citep[e.g.][]{yungelson2002,vanhaaften2012evo}. This assumption implies that these white dwarfs have cooled considerably by the time they eventually fill their Roche lobe, although tidal heating and irradiation before the onset of mass transfer may counteract this for a short time.\footnote{If the white dwarfs are warm, their radius is larger at a given mass \citep{bildsten2002,deloye2003} and therefore the orbital periods we find are strictly lower limits.} This leads to runaway mass loss on the dynamical timescale of the donor, followed by accretion of part of the disrupted white dwarf via a disk around the neutron star \citep[see e.g.][]{vandenheuvel1984bons,fryer1999,paschalidis2011,metzger2012}.
Furthermore, in systems with a donor mass larger than $\sim \! 0.38\ M_{\odot}$ \citep{yungelson2002,vanhaaften2012evo} (this value is only weakly sensitive to accretor mass) the accretor will be unable to eject enough transferred matter from the system by isotropic re-emission, where most arriving matter leaves the vicinity of the accretor in a fast, isotropic wind powered by accretion \citep{soberman1997,tauris1999}. This also leads to a merger. Therefore, systems that are unstable due to either a dynamical instability or insufficient isotropic re-emission have been removed from the sample. These instabilities only occur in systems with a white dwarf donor, because of the negative donor mass-radius exponent and the small donor size (hence, small orbit) at the onset of mass transfer. Dynamical instabilities may occur in systems with helium or main sequence donors if they have masses considerably higher than considered in this study \citep[see e.g.][]{pols1994}.

\begin{figure}
\resizebox{\hsize}{!}{\includegraphics{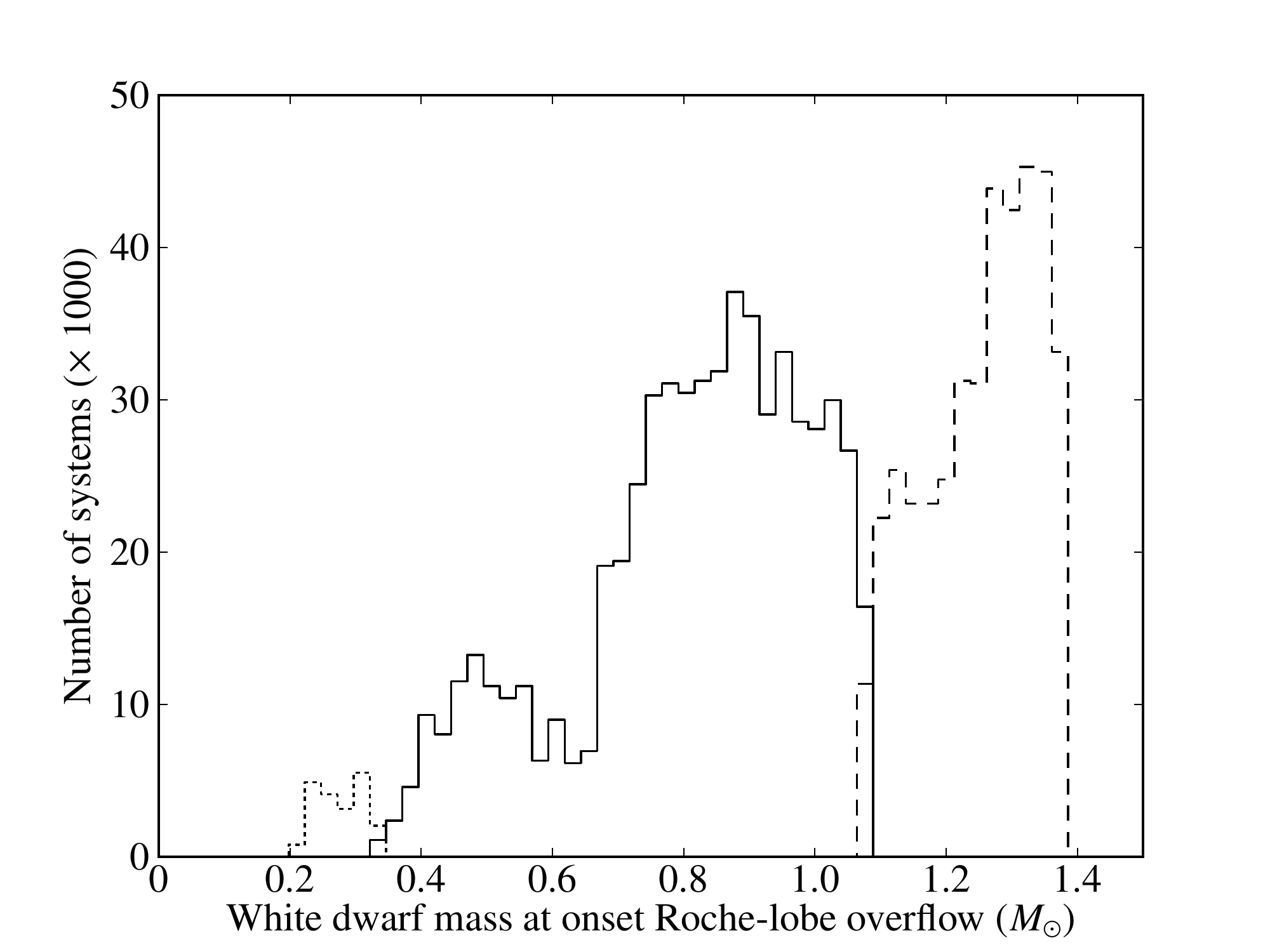}} 
\caption{Total time-integrated number of UCXBs with a helium (dotted line), carbon-oxygen (solid line) or oxygen-neon (dashed line) white dwarf donor at the onset of mass transfer to a neutron star (including merging systems).}
\label{fig:ucxb:hist_wd}
\end{figure}

In our simulation $97.4\%$ of all white dwarf systems have a donor with a mass higher than $0.38\ M_{\odot}$ and do not survive the onset of mass transfer. This includes $99.1\%$ of carbon-oxygen (solid line in Fig.~\ref{fig:ucxb:hist_wd}) and all oxygen-neon (dashed line in Fig.~\ref{fig:ucxb:hist_wd}) white dwarf systems. In about $80\%$ of the surviving white dwarf donor systems, the donor is a helium white dwarf (dotted line in Fig.~\ref{fig:ucxb:hist_wd}), in the remainder it is a carbon-oxygen white dwarf.
All surviving systems experienced the supernova explosion in the primary star and host a neutron star.

If a white dwarf donor with a mass higher than the $0.38\ M_{\odot}$ isotropic re-emission limit has a non-degenerate surface layer, the system may not merge immediately upon the onset of mass transfer, but it will once this layer has been lost.

\begin{figure}
\resizebox{\hsize}{!}{\includegraphics{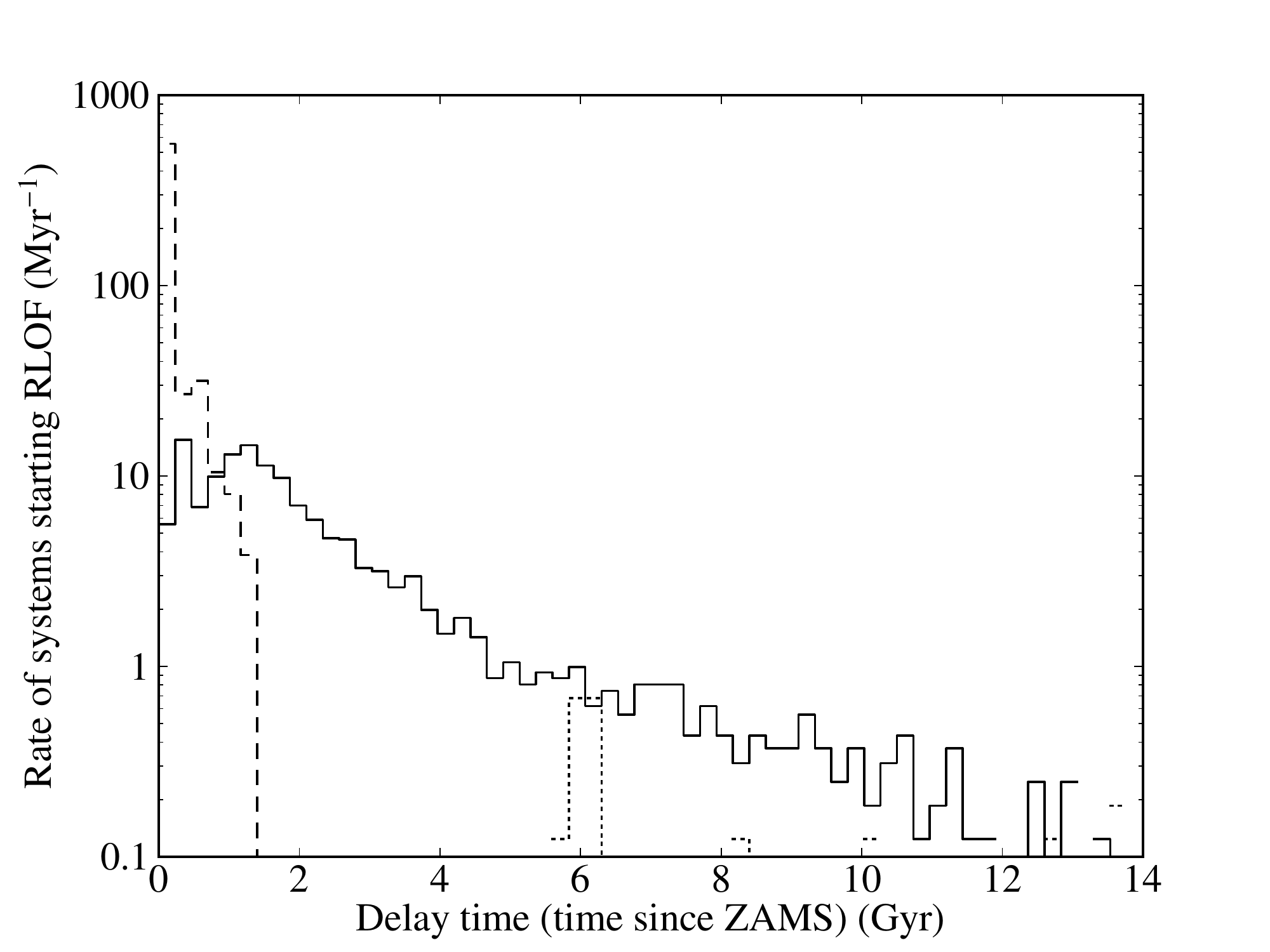}} 
\caption{Delay time distribution between the zero-age main sequence (ZAMS) and the onset of mass transfer to a neutron star for UCXBs with a white dwarf (excluding merging systems, solid line), helium star (dashed line) and main sequence (dotted line) donor.}
\label{fig:ucxb:delay}
\end{figure}

Two thirds of the white dwarf donor systems start to transfer mass to the neutron star within $2$ Gyr, but some systems take much longer (Fig.~\ref{fig:ucxb:delay}). This is the case for all white dwarf types. White dwarfs can take very long to start mass transfer depending on the width of the initial orbit, since the orbital decay of binaries consisting of a neutron star and a white dwarf is caused by gravitational wave radiation only.

\paragraph{Evolutionary tracks}

For each donor composition, the evolution after the stage described in Sect.~\ref{sect:ucxb:wdire} follows the tracks described in \citet{vanhaaften2012evo}.
Initially, the white dwarf donor has not yet cooled and therefore is larger than a zero-temperature white dwarf of the same mass. While the donor loses mass, its radius is held constant until it equals the zero-temperature radius of the same mass (this is justified by the rapid mass loss the donor initially experiences). From this point on, the zero-temperature radius \citep{zapolsky1969,rappaport1987} is used, which increases with further mass loss. The initial neutron star mass is taken to be $1.4\ M_{\odot}$ and its radius $12$ km \citep{guillot2011,steiner2013}. The evolution of UCXBs with degenerate donor stars is governed by angular momentum loss through gravitational wave radiation, which forces mass transfer via Roche-lobe overflow.

\subsubsection{Helium burning donor systems}

The supernova in low-mass helium burning star systems occurs in the primary, which has an initial mass $M \gtrsim 8\ M_{\odot}$. Most systems start out with an orbital period between $0.1$ and $100$ yr. Most of the helium stars form less than $500$ Myr after the zero-age main sequence when the secondaries, with initial masses of $2.3 - 5\ M_{\odot}$ experience hydrogen shell burning \citep{savonije1986} and lose their hydrogen envelope in case B mass transfer \citep{kippenhahn1967}. They fill their Roche lobes within another $\sim \! 200$ Myr, which is much earlier than UCXBs with white dwarf donors. In part this is due to the requirement that the helium star has not yet turned into a white dwarf before the onset of mass transfer (thereby disqualifying itself from the helium burning donor sample) which constrains the size of the initial orbit. We do not find systems with a black hole accretor.

\begin{figure}
\resizebox{\hsize}{!}{\includegraphics{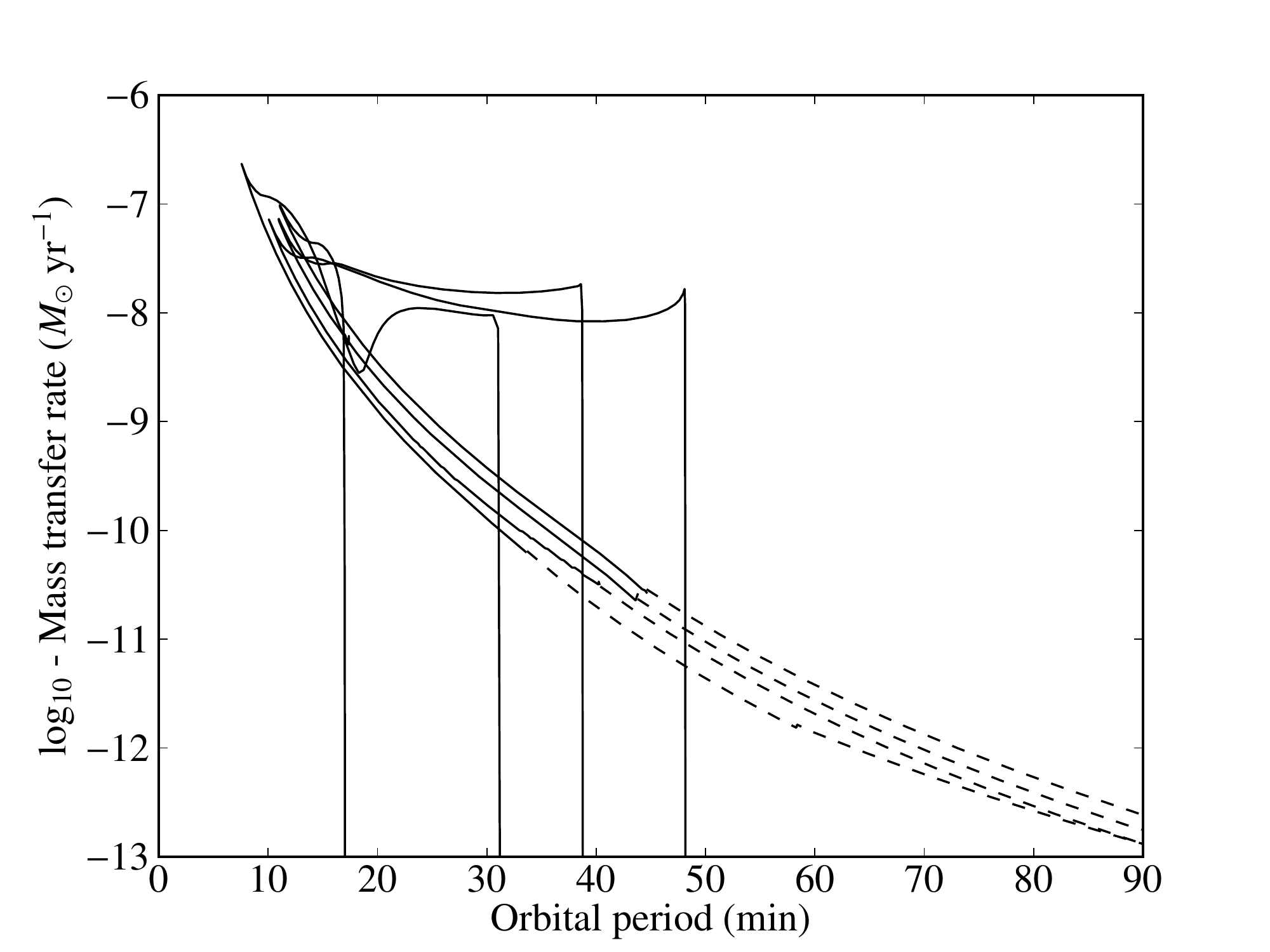}} 
\caption{Sample of helium star--neutron star UCXB tracks, for initial helium star mass range $0.35 - 1.0\ M_{\odot}$ and initial orbital period range $20 - 200$ min by \citet{nelemans2010} (solid lines), and our white dwarf donor track extensions (dashed lines).}
\label{fig:ucxb:hetracks}
\end{figure}

We have used stellar evolutionary tracks for systems with an $0.35 - 1.0\ M_{\odot}$ helium star and a $1.4\ M_{\odot}$ neutron star at initial orbital periods\footnote{The initial orbital period is the period directly after completion of the common-envelope phase that leaves behind the helium star \citep{yungelson2008}.} between $20$ and $200$ min \citep[][table in electronic article]{nelemans2010}, part of which is shown in Fig.~\ref{fig:ucxb:hetracks}. These tracks were made in the same way as the tracks for systems with white dwarf accretors in \citet{yungelson2008}. The donor metallicity $Z = 0.02$. Because the donors at the end of the tracks are degenerate, we have extended these tracks by making a smooth transition to the zero-temperature white dwarf evolution. The tracks describe the orbital period, mass transfer rate, donor mass, and core and surface compositions as a function of time.
In Fig.~\ref{fig:ucxb:hetracks}, the helium stars live up to $400$ Myr. After the onset of mass transfer (vertical part of the tracks), the orbits shrink until the period minimums, then expand towards the bottom right of the figure.
For each individual `zero-age' UCXB system produced by \textsf{SeBa}, the track that best matches its donor mass and orbital period has been used.

\subsubsection{Main sequence donor systems}

Main sequence donors have mostly evolved from $1.0 - 1.2\ M_{\odot}$ secondaries that started transferring mass after orbital decay due to magnetic braking \citep{sluys2005a}. After the magnetic field disappears (because the star becomes fully convective as a result of mass loss), gravitational wave radiation becomes the dominating angular momentum loss mechanism, continuing the orbital shrinking. In this scenario, the initial period and donor mass need to fall within relatively narrow ranges in order to sufficiently evolve the main sequence star. Moreover, the magnetic braking must be sufficiently efficient \citep{sluys2005b} which it probably is not \citep{queloz1998}. Depending on the extent of hydrogen depletion in the stellar center, systems can reach a minimum orbital period between $10$ and $80$ min, where $\sim \! 80$ min is the lower limit for hydrogen-rich donors \citep{paczynski1981cv}.

We have used stellar evolutionary tracks by \citet{sluys2005a} for binaries with an $0.7 - 1.5\ M_{\odot}$ main sequence star and a $1.4\ M_{\odot}$ neutron star at initial orbital periods between $0.50$ and $2.75$ days. The donor metallicity $Z = 0.01$. These tracks describe the orbital period and mass transfer rate as a function of time, as well as the core and surface compositions.

\subsection{Behavior of old UCXBs}
\label{sect:ucxb:old}

Figure~\ref{fig:ucxb:hetracks} suggests that once the donor has become degenerate, UCXBs `uneventfully' reach long orbital periods and very low mass transfer rates. This is probably not the case.
Instead, at low mass transfer rate a thermal-viscous instability in the accretion disk \citep{osaki1974,lasota2001} can cause UCXBs with a sufficiently low mass transfer rate to become transient \citep{deloye2003}. This usually implies that these systems are visible only during outbursts when the disk is in a hot state, which is only a small fraction of the time, and not during the quiescent state when the disk is cold and gaining mass.
Furthermore, due to accretion of angular momentum, a neutron star accretor in an UCXB can be recycled to a spin period between one and a few ms \citep{bisnovatyikogan1974,alpar1982,radhakrishnan1982}. Combined with a low mass transfer rate, the magnetosphere may transfer angular momentum from the neutron star to the accretion disk, thereby accelerating orbiting disk matter and counteracting accretion, known as the `propeller effect' \citep{davidson1973,illarionov1975}. As a result, the inner accretion disk, source of most X-ray radiation, can become disrupted by the magnetosphere. See \citet{vanhaaften2012evo} for more details on the thermal-viscous disk instability and propeller effect in UCXBs.
Finally, at low donor mass, high-energy radiation from the neutron star, the magnetosphere and the accretion disk may evaporate the donor, or detach it from its Roche lobe \citep{kluzniak1988,vandenheuvel1988,ruderman1989,rasio1989}. Hot, low-mass donors may suffer from a dynamical instability caused by a minimum value of their mass in the case of a constant core temperature \citep{bildsten2002}.

Each of these mechanisms can potentially diminish the visibility of UCXBs.
Because it is impossible to precisely quantify at which stage of the evolution (if at all) these mechanisms become important, and to what degree, we do not remove UCXBs from the sample, instead we will discuss the implications on the population of old UCXBs in the Discussion (Sect.~\ref{sect:ucxb:discussion}).

\subsection{Present-day population}
\label{sect:ucxb:method_present}

The present-day number and system parameters of UCXBs in the Galactic Bulge can be found by evaluating the evolutionary stage of all simulated systems at the present time. The most interesting parameters are the orbital period, mass transfer rate and surface composition, because these can be inferred from observations. The orbital periods of observed systems can be found from periodic modulations in the light curve or spectrum, although this is usually very difficult for UCXBs \citep[e.g.][]{nelemans2006,zand2007}. Since the transferred matter originates from the surface of the donor, the occurrence and relative abundance of elements in the donor can be inferred from X-ray \citep{schulz2001}, ultraviolet \citep{homer2002}, and optical \citep{nelemans2004} spectra, and more indirectly, type-I X-ray bursts \citep{zand2005}.
The mass transfer rate cannot be directly determined observationally. However, because the energy output of an X-ray binary is for a large part provided by the gravitational energy release of the accreted matter, the mass transfer rate strongly influences the luminosity of the system, which can be observed.

\subsubsection{Bolometric luminosity}
\label{sect:ucxb:xray}

We employ two methods of converting the modeled mass transfer rate to bolometric luminosity.

\begin{figure}
\resizebox{\hsize}{!}{\includegraphics{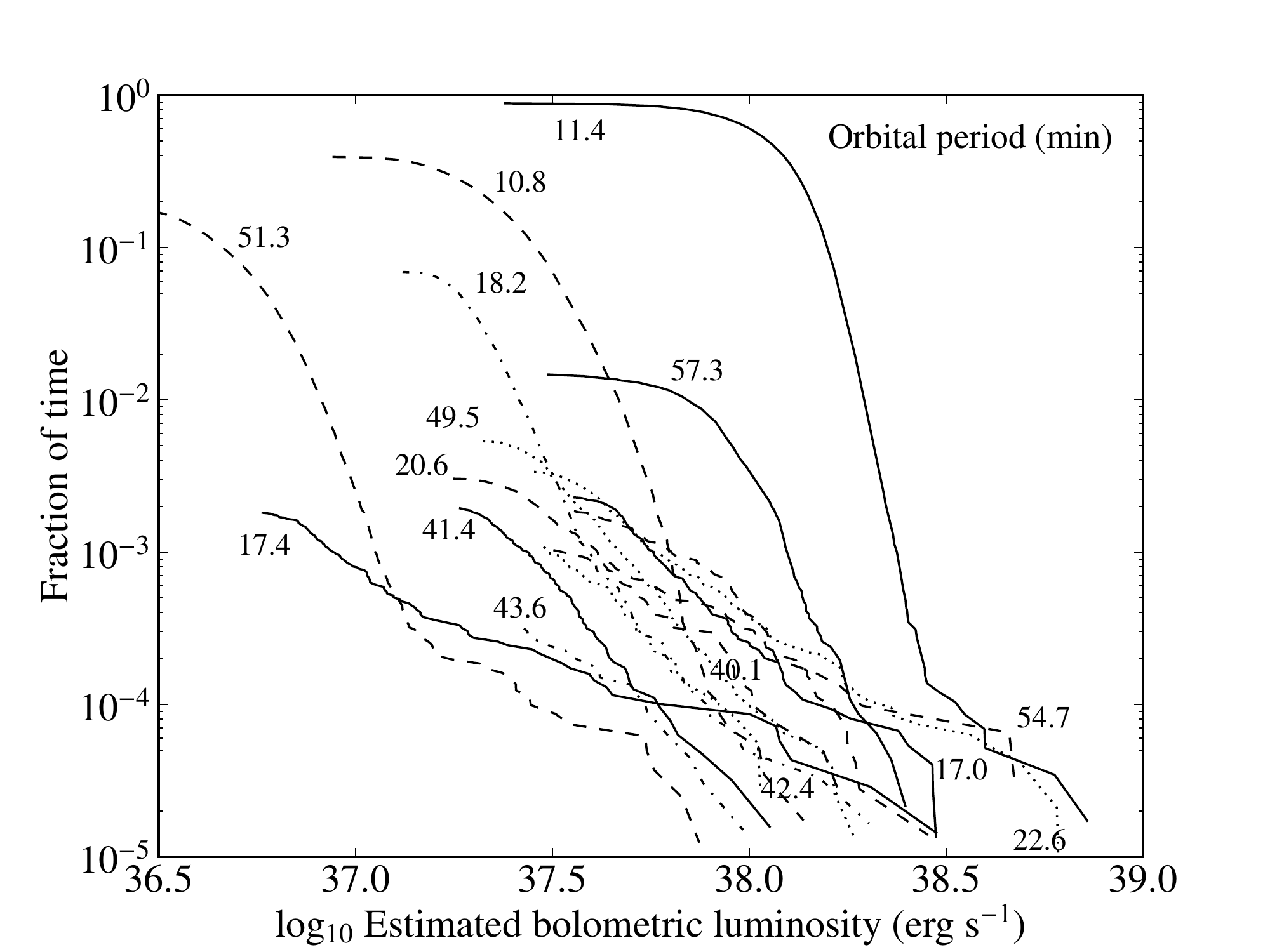}} 
\caption{UCXB variability: fraction of time that a source emits above a given luminosity for $14$ UCXBs, including two candidates with tentative orbital periods, adapted from \citet{vanhaaften2012xlum}. The numbers associated with the curves indicate the orbital periods in minutes.}
\label{fig:ucxb:signilum}
\end{figure}

An observational method is to match the modeled systems to real systems and assume that the modeled system behaves similar to the real system in terms of emission. We match a modeled UCXB to the real UCXB with the nearest orbital period. The relevant parameter of the emission behavior is the fraction of the time a source radiates at a given bolometric luminosity, measured over a sufficiently long timespan. We use $16$-year observations by the \emph{RXTE} ASM to determine this behavior for the $14$ known UCXBs (including two candidates) for which ASM data is available. Figure~\ref{fig:ucxb:signilum} shows this behavior for sources when they are visible well above the noise level \citep{vanhaaften2012xlum}. ASM X-ray luminosity was converted to bolometric luminosity using an estimate by \citet{zand2007}.
At a given time, the luminosity of an UCXB is randomly drawn from either the individual ASM observations that make up this time-luminosity curve, or (most of the time) from the faint-end extrapolations of the curves in Fig.~\ref{fig:ucxb:signilum} \citep{vanhaaften2012xlum}. These extrapolations are constructed in such a way that the average luminosity of the luminosity distribution is equal to the time-averaged luminosity of the source as observed by the ASM. The amount of time that a given source spends at a particular luminosity translates into the number of sources in a population at the same luminosity.

The second method, of a more theoretical nature, is to convert a system's modeled mass transfer rate to luminosity, using predictions by the disk instability model (Sect.~\ref{sect:ucxb:old}) in the case of long-period UCXBs. According to this model the mass transfer rate must exceed a critical value in order to be stable and the source to be persistent, i.e., visible at a relatively high luminosity (almost) all the time. A crude estimate for the critical mass transfer rate in the case of an irradiated disk is given by \citet{zand2007}, based on \citet{dubus1999,lasota2001,menou2002}
\begin{equation}
    \label{eq:ucxb:inst}
    \dot{M}_\mathrm{crit} \approx 5.3 \times 10^{-11}\ f \left( \frac{M_\mathrm{a}}{M_{\odot}} \right)^{0.3} \ \left( \frac{P_\mathrm{orb}}{\mathrm{h}} \right)^{1.4}\ \ M_{\odot}\ \mathrm{yr}^{-1}
\end{equation}
with $M_\mathrm{a}$ the accretor mass, $P_\mathrm{orb}$ the orbital period, and $f$ is a factor accounting for the disk composition; $f \approx 1$ for carbon-oxygen disks and $f \approx 6$ for helium disks.

When the time-averaged mass transfer rate exceeds the critical value, the bolometric luminosity $L$ is assumed to be constant at
\begin{equation}
    \label{eq:ucxb:lum}
    L = \frac{GM_\mathrm{a}\dot{M}_\mathrm{a}}{2 R_\mathrm{a}},
\end{equation}
with $G$ the gravitational constant and $R_\mathrm{a}$ the accretor radius.

Sources with a time-averaged mass transfer rate below the critical value are assumed to be visible only during outburst stages. The predictions by the thermal-viscous disk instability model regarding the degree of variability of sources is supported by observations \citep{vanparadijs1996,ramsay2012,coriat2012}. The duty cycle (fraction of the time the source is in outburst) is
\begin{equation}
    \label{eq:ucxb:dutycycle}
    \mathrm{DC} = \frac{L_\mathrm{avg}}{L_\mathrm{outburst}},
\end{equation}
where $L_\mathrm{avg}$ is the time-averaged luminosity based on the theoretical mass transfer rate, (Eq.~\ref{eq:ucxb:lum}), and $L_\mathrm{outburst}$ is the luminosity during outburst, derived by \citet{lasotaetal2008}
\begin{equation}
    \label{eq:ucxb:lburst_he}
    L_\mathrm{outburst} \approx 3.5 \times 10^{37}\ \left( \frac{P_\mathrm{orb}}{\mathrm{h}} \right)^{1.67}\ \mbox{erg s}^{-1},
\end{equation}
which is consistent with observations of outbursts in UCXBs \citep[e.g.][]{wu2010}.
The period of this cycle is not relevant here. We neglect the decay in the light curve after an outburst. Furthermore, we do not predict the luminosity of systems that are in quiescence, which in fact has been assumed to be zero in the above method.

Both methods have advantages and shortcomings.
The ASM observations have a rather high lower limit in converted bolometric luminosity, of $\sim \! 10^{37}\ \mbox{erg s}^{-1}$ at $8.3$ kpc, the estimated distance to the Galactic Center \citep{gillessen2009}. Below this luminosity, we have to rely on extrapolations. Also, variability on a timescale longer than $16$ yr cannot have been observed by the ASM. On the other hand, the data do not strongly rely on uncertainties in models, as is the case for the disk instability model method.
The ASM data show that UCXBs with a similar orbital period can behave rather differently, for instance \object{XTE J1751--305} ($42.4$ min orbital period) and \object{XTE J0929--314} ($43.6$ min), or \object{4U 0513--40} ($17.0$ min) and \object{2S 0918--549} ($17.4$ min).
Equation~(\ref{eq:ucxb:inst}), however, shows that in the disk instability model the critical mass transfer rate and hence luminosity is largely determined by the orbital period and composition.
A longer orbital period does not automatically imply a lower luminosity, as evidenced by \object{M 15 X-2} ($20.6$ min orbital period) and \object{4U 1916--05} ($49.5$ min).

In general, the ASM data show that a clear distinction between persistent and transient behavior is not justified \citep{vanhaaften2012xlum}. Almost all systems are visible above the ASM detection limit only sporadically. Still, short orbital period systems are typically visible more often at a given luminosity, i.e., they have a (slightly) higher time-averaged luminosity. Even though the available sample is small, individual unusual behavior is expected to partially cancel out for the population as a whole, because some modeled UCXBs will be matched to a real UCXB that is brighter than typical for its orbital period, while others will be matched to one that is fainter than typical.

\subsubsection{Donor composition}
\label{sect:ucxb:surf}

The donor surface compositions of the modeled UCXBs are predicted using the helium-star donor and main sequence donor tracks, as well as the white dwarf types from the population synthesis model. These predicted compositions can be compared to observations of real systems, in the Bulge and elsewhere.
Donors that start mass transfer as a white dwarf can be helium and carbon-oxygen white dwarfs (Sect.~\ref{sect:ucxb:wdire}). In the latter case we assume $30\%$ carbon and $70\%$ oxygen by mass, based on the most common eventual compositions in the helium burning donors (Sect.~\ref{sect:ucxb:donsurf}). Donors that start mass transfer as a helium burning star can also produce helium-carbon-oxygen donors due to an interrupted helium burning stage. For the subsequent tracks for these systems, we use the mass-radius relation for degenerate donors composed of a mixture of $60\%$ helium, $30\%$ carbon and $10\%$ oxygen, a choice based on the dominant tracks by \citet{nelemans2010} as will be discussed in Sect.~\ref{sect:ucxb:donsurf}. We note that the degenerate tracks are not very sensitive to the composition (as long as there is no hydrogen), so these simplifications are justified.
Matter processed in the CNO cycle has a high nitrogen-to-carbon abundance ratio, whereas helium burning converts this to a low ratio. Consequently, the nitrogen-to-oxygen ratio is a good test for the formation channel because it can discriminate between a history as a helium white dwarf donor or a helium burning donor \citep{nelemans2010}.

Based on the overview in \citet{vanhaaften2012xlum}, the compositions of observed UCXBs can be summarized as being roughly equally distributed over helium and carbon-oxygen compositions. There is no clear dependency on the orbital period, although helium composition may be more common among systems with a long orbital period ($\gtrsim 40$ min) \citep{vanhaaften2012xlum}.
The surface composition of very low-mass donors corresponds to the (inner) core composition of the object before it started transferring mass.

\section{Results}
\label{sect:ucxb:results}

\subsection{Birth rates and total number of systems}

\begin{figure}
\resizebox{\hsize}{!}{\includegraphics{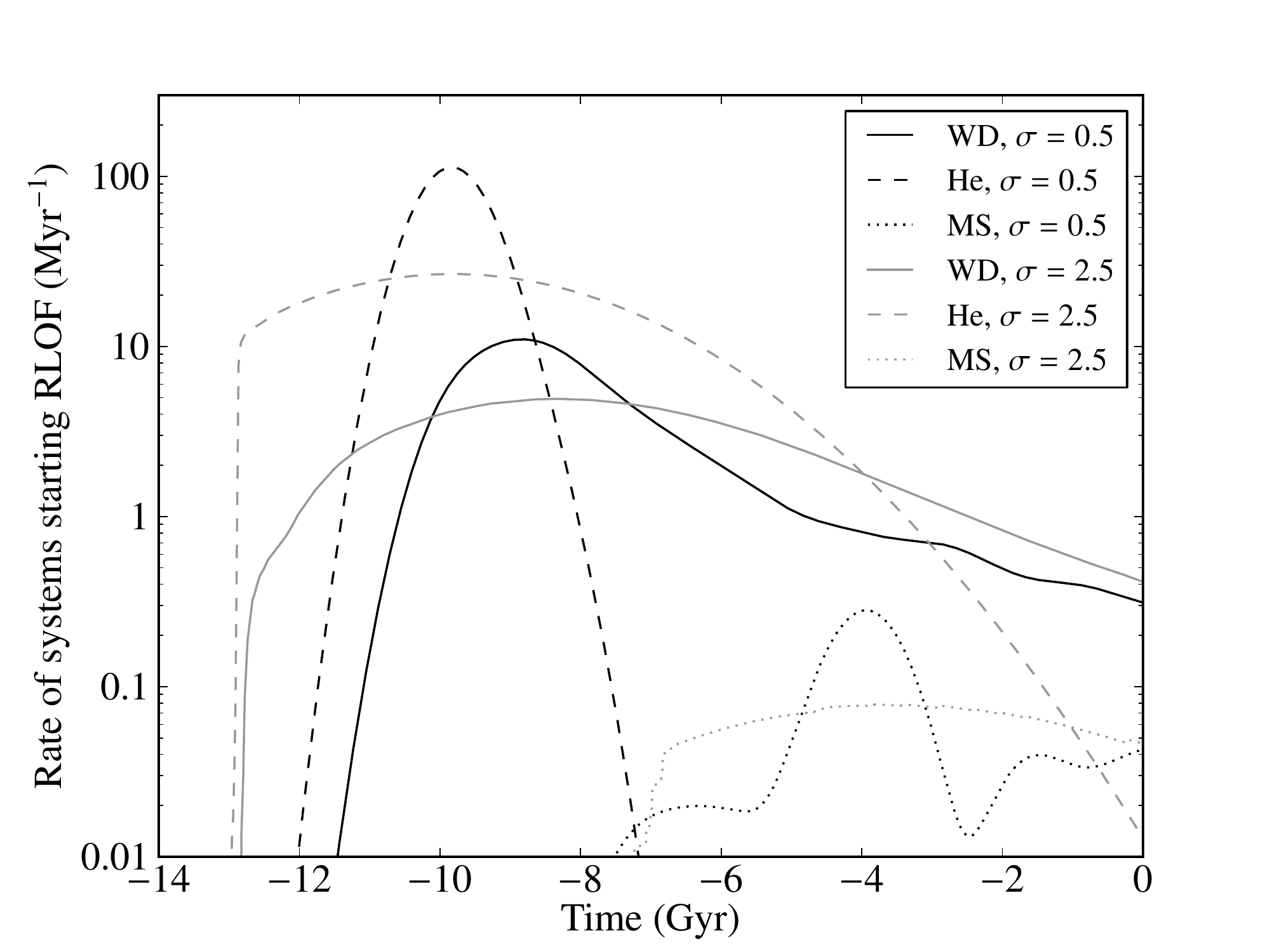}} 
\caption{Birth rate of systems reaching Roche-lobe overflow against time for UCXBs from the white dwarf donor channel (solid lines), helium-star donor channel (dashed lines) and main sequence donor channel (dotted lines). Time $= 0$ corresponds to the present. Black lines correspond to a star formation history width $\sigma = 0.5$ Gyr, gray lines to $\sigma = 2.5$ Gyr.}
\label{fig:ucxb:brate}
\end{figure}

Convolving the star formation history (Fig.~\ref{fig:ucxb:sfh}) with the delay times of the onset of mass transfer (Fig.~\ref{fig:ucxb:delay}) yields the birth rate distributions, shown in Fig.~\ref{fig:ucxb:brate} for burst-like (black) and extended (gray) star formation epochs. Except for part of the white dwarf channel and the main sequence channel, the delay times are much shorter than the age of the Bulge (Sect.~\ref{sect:ucxb:sfhini}). In the case of $\sigma = 0.5$  Gyr, $98\%$ of the white dwarf donor systems have started Roche-lobe overflow before the present, whereas $100\%$ of the helium-star donor and $84\%$ of the main sequence donor systems have. In the case of $\sigma = 2.5$, these percentages are $97\%$, $100\%$ and $77\%$, respectively. The main distinguishing feature between the three classes (Sect.~\ref{sect:ucxb:formation}) is the most recent time at which mass transfer can begin. Initially wide systems from the white dwarf donor channel can still start mass transfer at the present, whereas main sequence donor systems and especially helium-star donor systems cannot, unless they have formed relatively recently (star formation history width $\sigma = 2.5$ Gyr). The rate of helium burning donor systems reaching Roche-lobe overflow closely follows the star formation history.

Upon the onset of mass loss, the donor radius increases immediately for fully degenerate white dwarf donors, and after approximately $100$ Myr for helium burning donors, once the donor has become sufficiently degenerate following the extinction of nuclear fusion (this happens some time after the period minimum). The orbital period decreases in the case of helium burning or main sequence donors, whereas the period increases with mass loss for systems with degenerate donors (in each channel). If main sequence donor systems become ultracompact, this typically happens $\sim \! 3$ Gyr after the onset of mass transfer, and mass transfer starts after $2 - 6$ Gyr after the formation of the binary \citep{sluys2005a}.

\begin{figure}
\resizebox{\hsize}{!}{\includegraphics{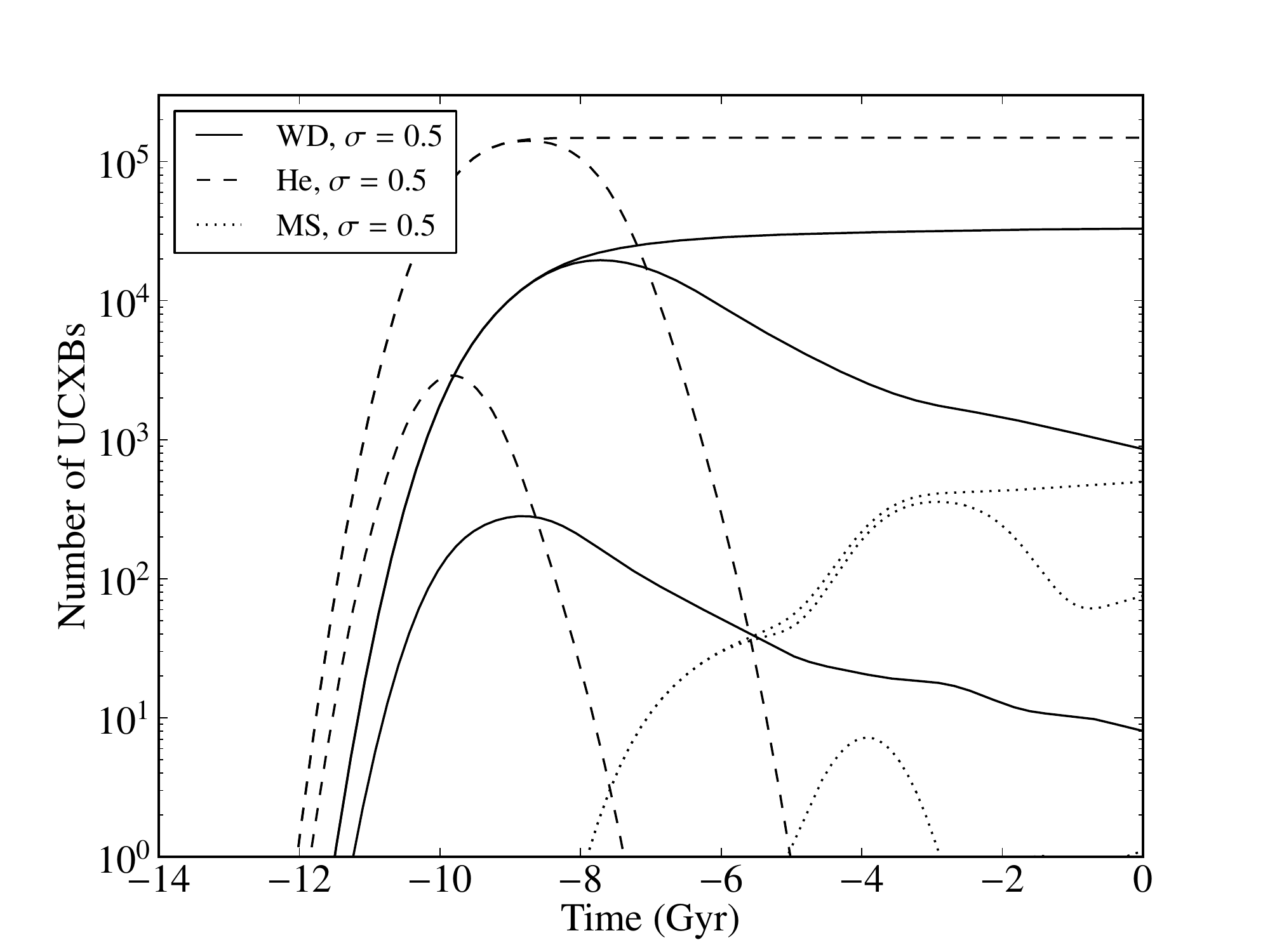}} 
\caption{Number of UCXBs from the white dwarf donor channel (solid lines), helium-star donor channel (dashed lines) and main sequence donor channel (dotted lines). Time $= 0$ corresponds to the present. For each line style, the three lines represent the full population (upper) and the systems with orbital periods shorter than $60$ min (middle) and shorter than $20$ min (lower). A star formation history width $\sigma = 0.5$ Gyr has been used.}
\label{fig:ucxb:history}
\end{figure}

The total number of UCXBs with a white dwarf or helium-star donor, shown as the solid and dashed lines in Fig.~\ref{fig:ucxb:history}, initially follows the star formation rate and later on approaches an upper limit as star formation slows down. The number of systems below a given orbital period initially resembles the instantaneous star formation rate more closely for short orbital periods. After the peak in star formation rate, the number of systems below a given period keeps increasing as long as more new systems form than old systems are removed from the given sample due to their increasing orbital periods. The numbers of systems from the three different classes decline at different rates corresponding to their respective recent birth rates (Fig.~\ref{fig:ucxb:brate}).

\subsection{Present-day population}
\label{sect:ucxb:totalpop}

\begin{figure}
\resizebox{\hsize}{!}{\includegraphics{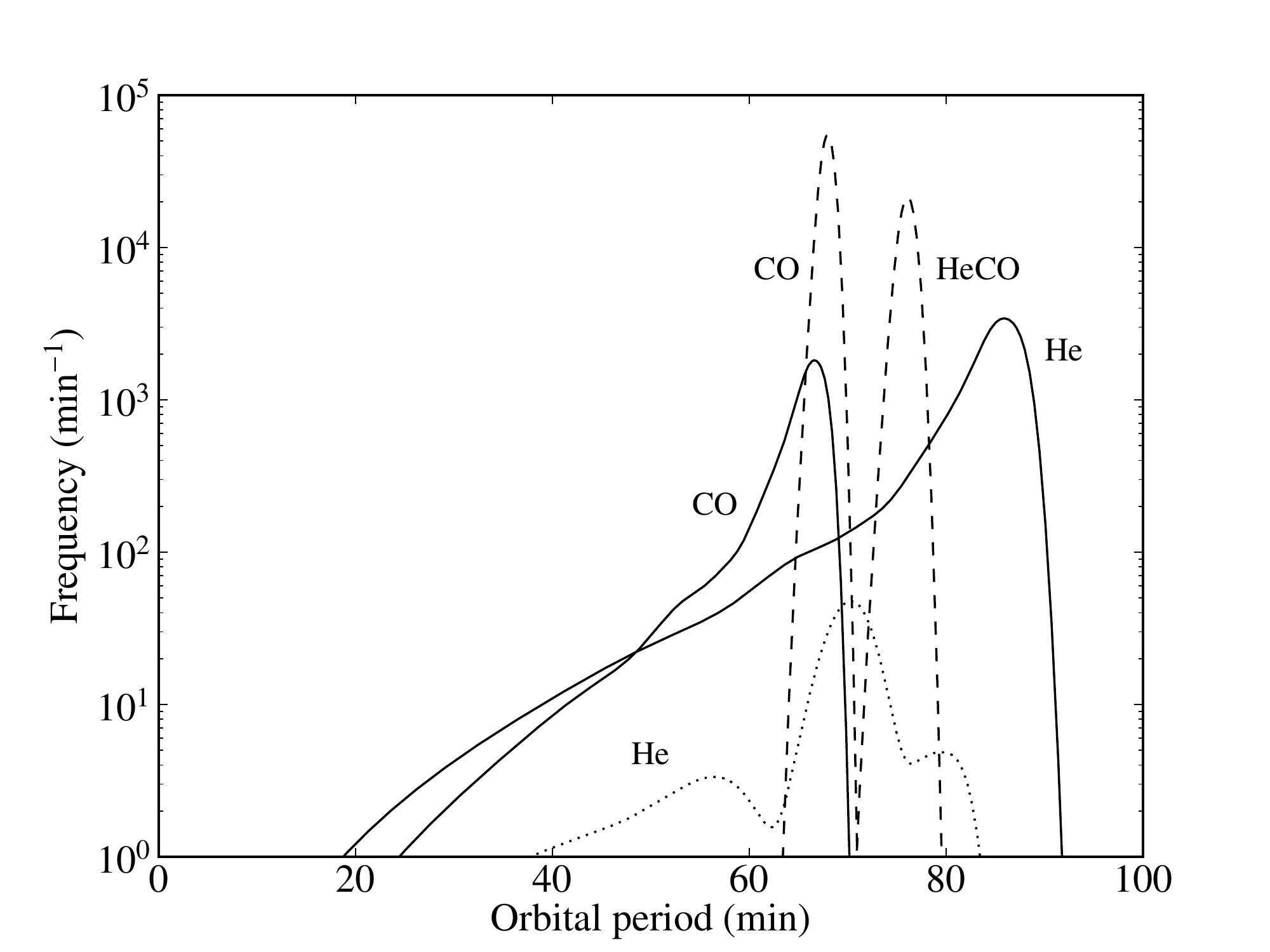}} 
\caption{Present-day orbital period distribution for UCXBs from the white dwarf donor channel (solid lines), helium-star donor channel (dashed lines) and main sequence donor channel (dotted lines). The elements next to the lines indicate the most abundant element(s) at the surface of the donor, hence in the transferred matter. The star formation history has width $\sigma = 0.5$ Gyr.}
\label{fig:ucxb:presentpop}
\end{figure}

While the evolution of the population is interesting in itself, the population today can be used to validate the results.
In the case of a star formation history distribution width $\sigma = 0.5$ Gyr, shown in Fig.~\ref{fig:ucxb:presentpop}, most systems at the present are old, and have expanded to an orbital period of $\sim \! 80$ min. Since evolution slows down at longer periods, systems tend to `pile up'.\footnote{For a donor mass-radius exponent $\zeta$, the number of UCXBs at a given period $N \propto P_\mathrm{orb}^{(11/3 - 5\zeta)/(1 - 3\zeta)}$ \citep{deloye2003}.} Differences in donor composition lead to different present-day orbital periods. This is the case even among hydrogen-deficient compositions because during most of the evolution, the donor mass is low enough for Coulomb physics to be important to the stellar structure, or even dominating degeneracy pressure. Coulomb interactions cause a donor that is composed of `heavy' elements such as carbon and oxygen to have a smaller radius than donors with lighter composition, such as helium, of the same mass \citep{zapolsky1969}. A larger donor radius (at each mass) results in a longer orbital period at each mass, but also at each age (because less time is spent at a given orbital period).\footnote{However, more time is spent at a given donor mass.}

\begin{figure}
\resizebox{\hsize}{!}{\includegraphics{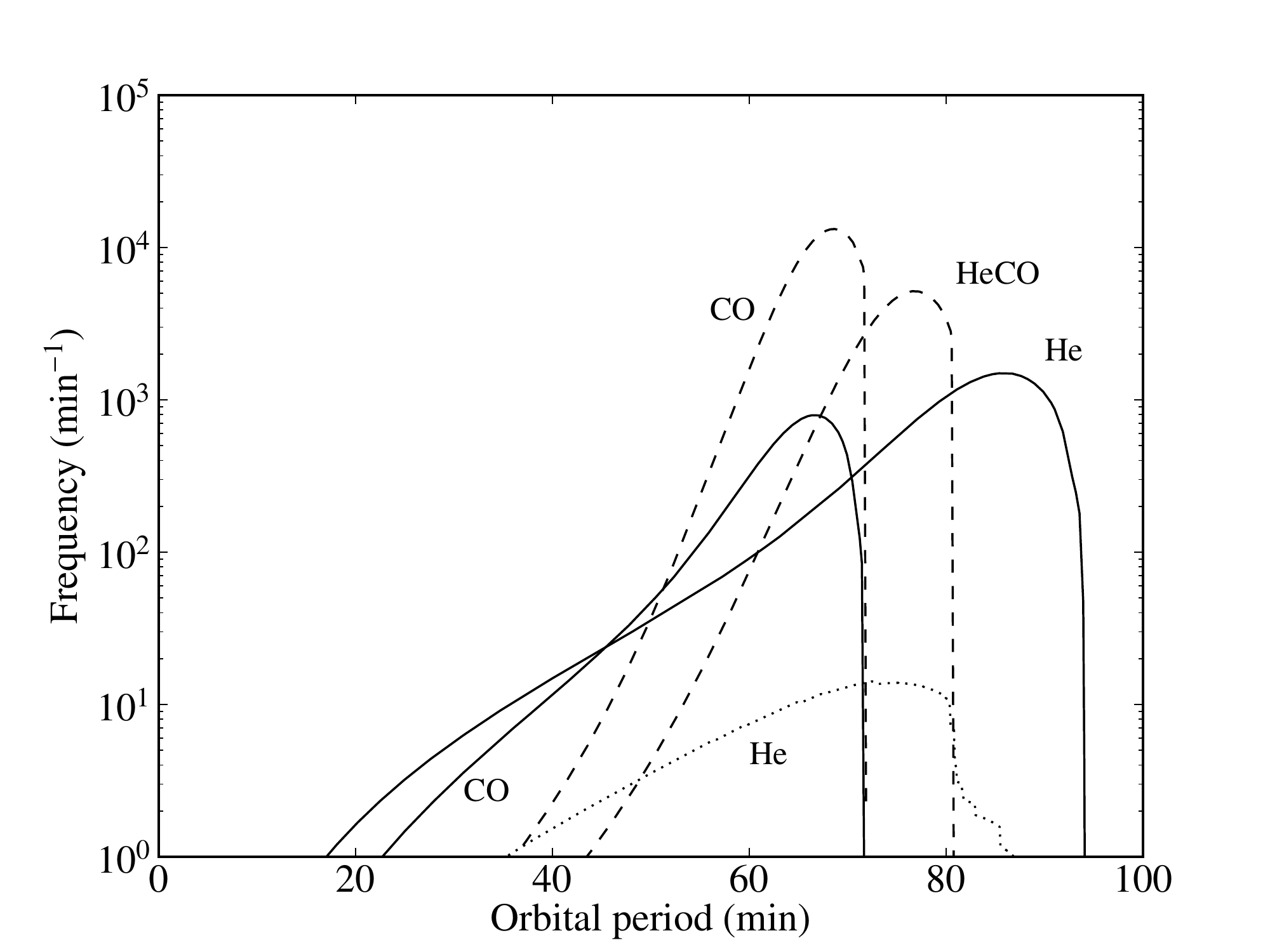}} 
\caption{Same as Fig.~\ref{fig:ucxb:presentpop} except $\sigma = 2.5$ Gyr.}
\label{fig:ucxb:presentpop2}
\end{figure}

For $\sigma = 0.5$ Gyr, all UCXBs with orbital periods shorter than $1$ h started Roche-lobe overflow from a white dwarf donor (Fig.~\ref{fig:ucxb:presentpop}), long after the formation of the system. Most of these systems host a helium white dwarf. The main sequence channel contributes a negligible number of UCXBs and can only be distinguishable (in principle) via donor compositions.
For $\sigma = 2.5$ Gyr, shown in Fig.~\ref{fig:ucxb:presentpop2}, there is also recent star formation. This produces a population of young UCXBs that descended from helium burning donor systems (or still have a helium burning donor), with orbital periods shorter than $1$ h. The steep cut-off at the long-period end of several curves is due to the assumption that star formation suddenly starts $13$ Gyr before present.

Combining all donor compositions, the result is a current $\sim \! 1.9 \times 10^{5}$ population of UCXBs, mostly at long orbital period ($60 - 90$ min). The total number of UCXBs in each class is $3.5 \times 10^{4}$ ($18\%$) with white dwarf donors, $1.56 \times 10^{5}$ ($81\%$) with helium-star donors, and $5.1 \times 10^{2}$ ($0.3\%$) with main sequence donors.
The number of modeled systems with orbital periods shorter than $60$ min is $1.5 \times 10^{3}$ for $\sigma = 0.5$ Gyr, and $7.4 \times 10^{3}$ for $\sigma = 2.5$ Gyr ($0.8\%$ and $3.8\%$ of the population, respectively). We note that these numbers are rather sensitive to assumptions in the model, and could be lower by an order of magnitude, as will be discussed in Sect.~\ref{sect:ucxb:discussion}.

\subsection{Observable population}
\label{sect:ucxb:resobs}

As described in Sect.~\ref{sect:ucxb:xray}, in order to determine what we can observe at high luminosity we have to convert the modeled population to luminosities.

\subsubsection{\emph{RXTE} All-Sky Monitor}
\label{sect:ucxb:asm}

\begin{figure}
\resizebox{\hsize}{!}{\includegraphics{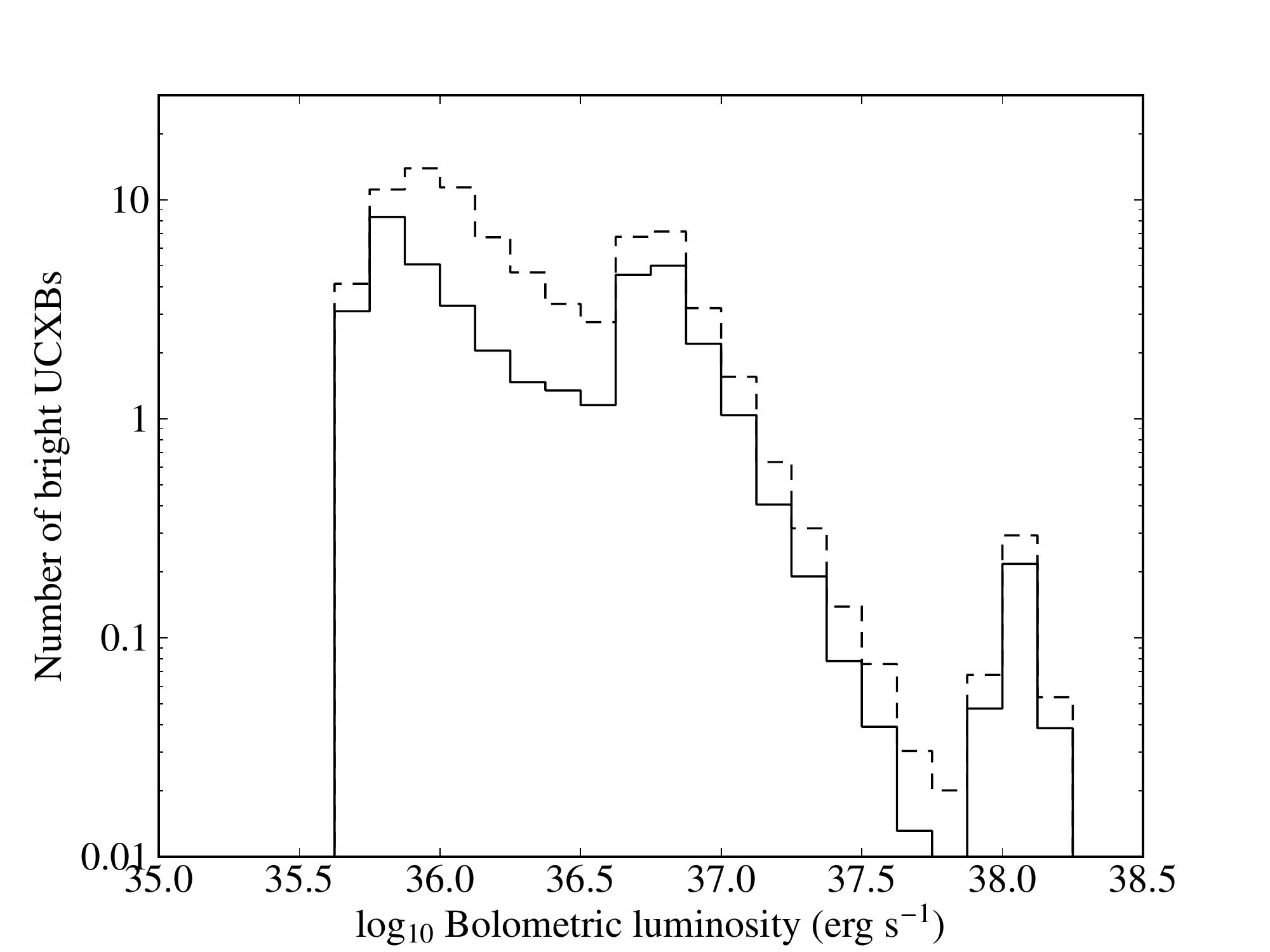}} 
\caption{Present-day luminosity distribution of UCXBs in the Bulge based on \emph{Rossi XTE} All-Sky Monitor observations, after incorporating the accelerated evolution of the systems (Sect.~\ref{sect:ucxb:asm}). The star formation history has width $\sigma = 0.5$ Gyr (solid line) and $\sigma = 2.5$ Gyr (dashed line).}
\label{fig:ucxb:presentlum_asm}
\end{figure}

In the first method, we apply the observations of known UCXBs by the \emph{RXTE} ASM (Fig.~\ref{fig:ucxb:signilum}, Sect.~\ref{sect:ucxb:xray}) to the modeled population (Figs.~\ref{fig:ucxb:presentpop} and \ref{fig:ucxb:presentpop2}, Sect.~\ref{sect:ucxb:totalpop}). Modeled UCXBs with an orbital period longer than $60$ min are left out because of the absence of known real systems with such periods (i.e., they are assumed never to reach luminosities above $\sim \! 10^{34}\ \mbox{erg s}^{-1}$). The time-averaged luminosity of most UCXBs with orbital periods longer than $40$ min is approximately two orders of magnitude higher than expected from the gravitational-wave model \citep{vanhaaften2012xlum}. This implies that either the observed sources are atypically bright, or that they show normal behavior, but evolve much faster than if driven only by gravitational wave losses (the implications will be discussed in Sect.~\ref{sect:ucxb:disclong}). In each case, due to energy conservation, we need to reduce the number of bright sources at each orbital period by a factor that corresponds to the ratio between the gravitational-wave luminosity and the actual observed luminosity, given by \citet[][their Fig.~3]{vanhaaften2012xlum}. Figure~\ref{fig:ucxb:presentlum_asm} shows the resulting number of bright UCXBs predicted by the ASM data.

For star formation history width $\sigma = 0.5$ Gyr, about $40$ systems are expected to be visible as bright sources at a given time. For $\sigma = 2.5$ Gyr, this number is larger, $\sim \! 80$, because recent star formation causes more young systems to exist, which have not yet reached orbital periods of $60$ min.
The cut-off at the faint end of the histogram is an artifact of the assumed linear extrapolation to the faint behavior. This also results in a relatively high minimum luminosity. In reality, especially sources with orbital periods longer than $\sim \! 40$ min are expected to be very faint (i.e., much fainter than suggested by a linear extrapolation) at least some fraction of the time, which means the cumulative luminosity distribution flattens at faint luminosities, causing a tail at the low-luminosity end of Fig.~\ref{fig:ucxb:presentlum_asm}. The distribution decreases with increasing luminosity in a similar way as the luminosity distribution of the representative individual observed UCXBs (Fig.~\ref{fig:ucxb:signilum}).

\subsubsection{Disk instability model}
\label{sect:ucxb:resdim}

\begin{figure}
\resizebox{\hsize}{!}{\includegraphics{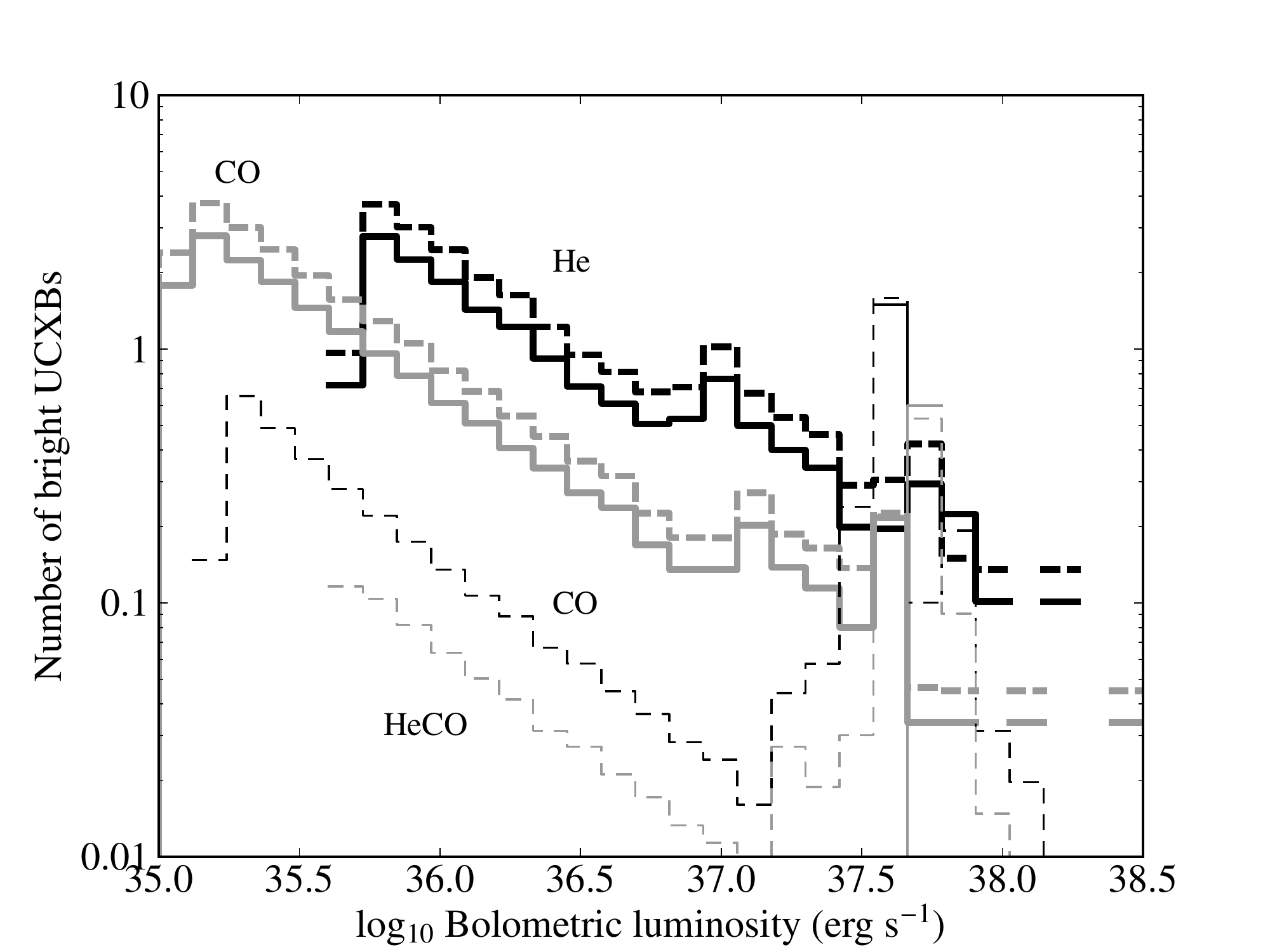}} 
\caption{Present-day luminosity distribution of UCXBs in the Bulge based on the disk instability model (Sect.~\ref{sect:ucxb:xray}). Line color distinguishes between helium (thick black lines), carbon-oxygen (thick gray and thin black lines) and helium-carbon-oxygen (thin gray lines) donor compositions. The thick lines correspond to the white dwarf donor channel, the thin lines to the helium burning channel. The star formation history has width $\sigma = 0.5$ Gyr (solid lines) and $\sigma = 2.5$ Gyr (dashed lines).}
\label{fig:ucxb:presentlum_dim}
\end{figure}

\begin{figure}
\resizebox{\hsize}{!}{\includegraphics{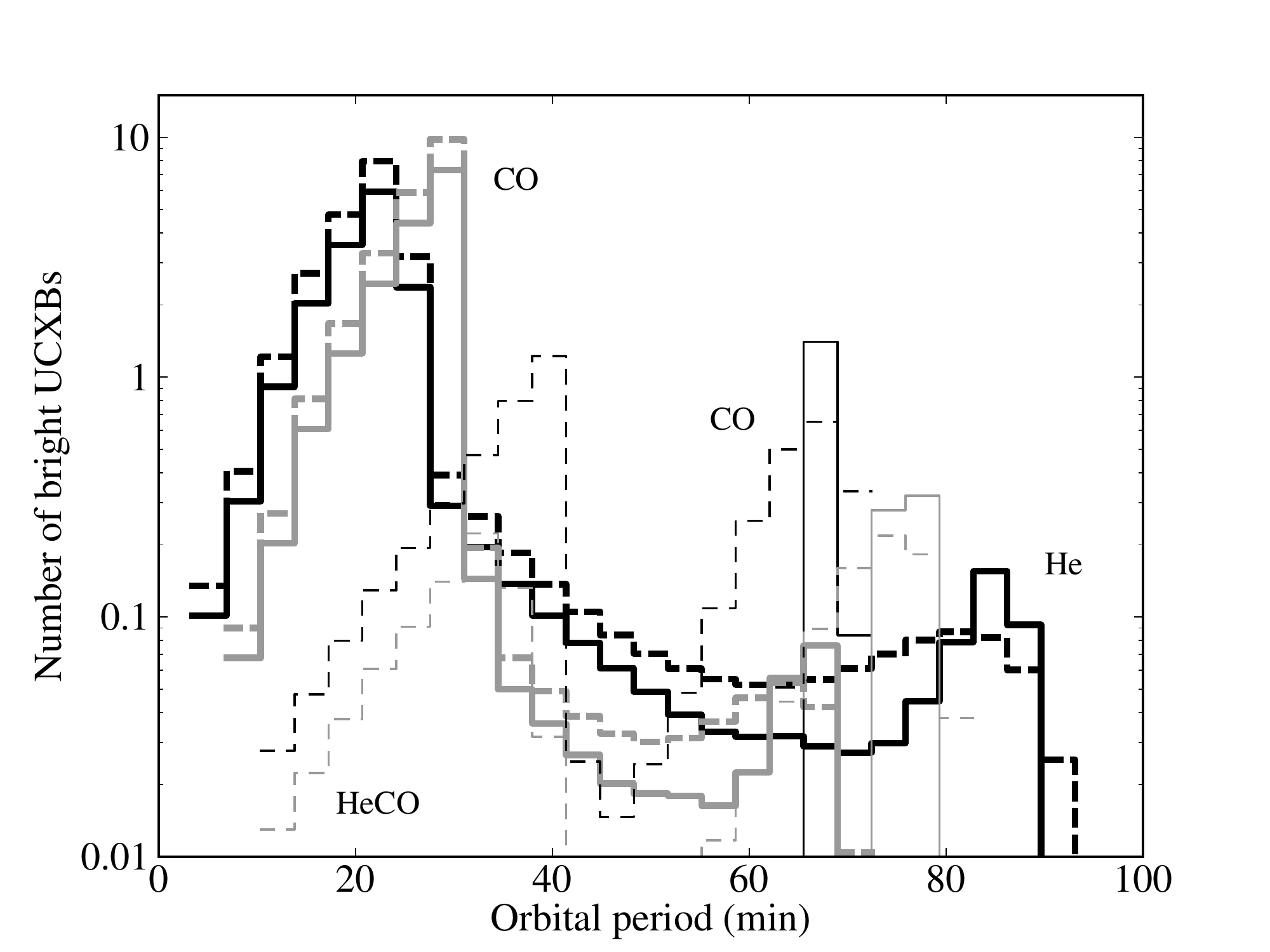}} 
\caption{Orbital period distribution of the predicted bright population of UCXBs in the Bulge at the present based on the disk instability model. For details see Fig.~\ref{fig:ucxb:presentlum_dim}.}
\label{fig:ucxb:presentlum_dim2}
\end{figure}

The second method relies on converting theoretical mass transfer rates to luminosities using the disk instability model (Sect.~\ref{sect:ucxb:xray}), the result of which is shown in Figs.~\ref{fig:ucxb:presentlum_dim} and \ref{fig:ucxb:presentlum_dim2}. The total number of bright ($\gtrsim \! 10^{35}\ \mbox{erg s}^{-1}$) sources (either persistent or in outburst) is $34$ for $\sigma = 0.5$ Gyr and $51$ for $\sigma = 2.5$ Gyr. The vast majority of these are persistent (short-period) sources, and therefore the number is larger in the case of recent star formation. For the same reason, the white dwarf donor channel (Sect.~\ref{sect:ucxb:wdire}) dominates this population, and most should have helium or carbon-oxygen donors. The helium burning channel is expected to contribute at most a couple of bright sources, in outburst at long orbital period ($60 - 80$ min).

Figure~\ref{fig:ucxb:presentlum_dim} shows the luminosity distribution of bright sources. UCXBs with a luminosity between $\sim \! 10^{36-37}\ \mbox{erg s}^{-1}$ are the ones with the shortest orbital periods, below $\sim \! 20$ min. From here, the number of sources at a given luminosity increases towards fainter luminosities because these sources have longer period derivatives and lower time-averaged mass transfer rates, until a sharp cut-off defined by the longest orbital period at which sources are still considered persistent. Carbon-oxygen dominated sources are persistent to longer periods and lower luminosities because accretion disks composed of carbon-oxygen are more stable than helium dominated disks, see Eq.~(\ref{eq:ucxb:inst}). For $\sigma = 0.5$ Gyr the number of short-period, persistent, systems is negligible.
At longer periods ($\gtrsim 40$ min), the duty cycle (Eq.~\ref{eq:ucxb:dutycycle}) determines the number of sources in outburst. The duty cycle decreases below $10^{-4}$ at orbital periods longer than $60 - 70$ min, depending on donor composition. During their rare outbursts, sources are temporarily bright at $\sim \! 10^{37-38}\ \mbox{erg s}^{-1}$, as follows from Eq.~(\ref{eq:ucxb:lburst_he}). Their peak luminosities are used in Fig.~\ref{fig:ucxb:presentlum_dim}.

Two peaks can be distinguished in the lines representing the white dwarf channel in Fig.~\ref{fig:ucxb:presentlum_dim} (thick lines). The peak at $\sim \! 10^{37}\ \mbox{erg s}^{-1}$ consists of systems with orbital periods just longer than the critical period because these still have a relatively high duty cycle. The second peak at $\sim \! 10^{37.6}\ \mbox{erg s}^{-1}$ consists of long-period systems because these are very numerous and distributed over a relatively narrow interval of orbital periods. The duty cycle at a given orbital period is higher for helium burning systems owing to their larger size and, because their average density is set by the orbital period, correspondingly larger mass. Hence, their time-averaged mass transfer rate at a given orbital period is also higher.

In Fig.~\ref{fig:ucxb:presentlum_dim2} the orbital period distribution is shown for the same population of bright sources as in Fig.~\ref{fig:ucxb:presentlum_dim}. The jumps of these distributions correspond to the respective low-luminosity ends of the distribution in Fig.~\ref{fig:ucxb:presentlum_dim}. The cut-off period of persistent sources (at $30-40$ min) lies at a longer orbital period for the systems with a helium burning donor origin compared with systems with a white dwarf origin. The reason is that these donors have a higher temperature than originally white dwarf donors, and therefore the time-averaged mass transfer rate is higher at the same period. This causes the disk to remain stable (and the sources to be persistent) up to a longer period. Again we see that carbon-oxygen donor systems are persistent up to a longer orbital period than helium-dominated donor systems. Transient systems with orbital periods longer than $\sim \! 40$ min are rarely in outburst and at most a handful have a high luminosity at a given time.

\subsection{Donor surface composition}
\label{sect:ucxb:donsurf}

\begin{figure}
\resizebox{\hsize}{!}{\includegraphics{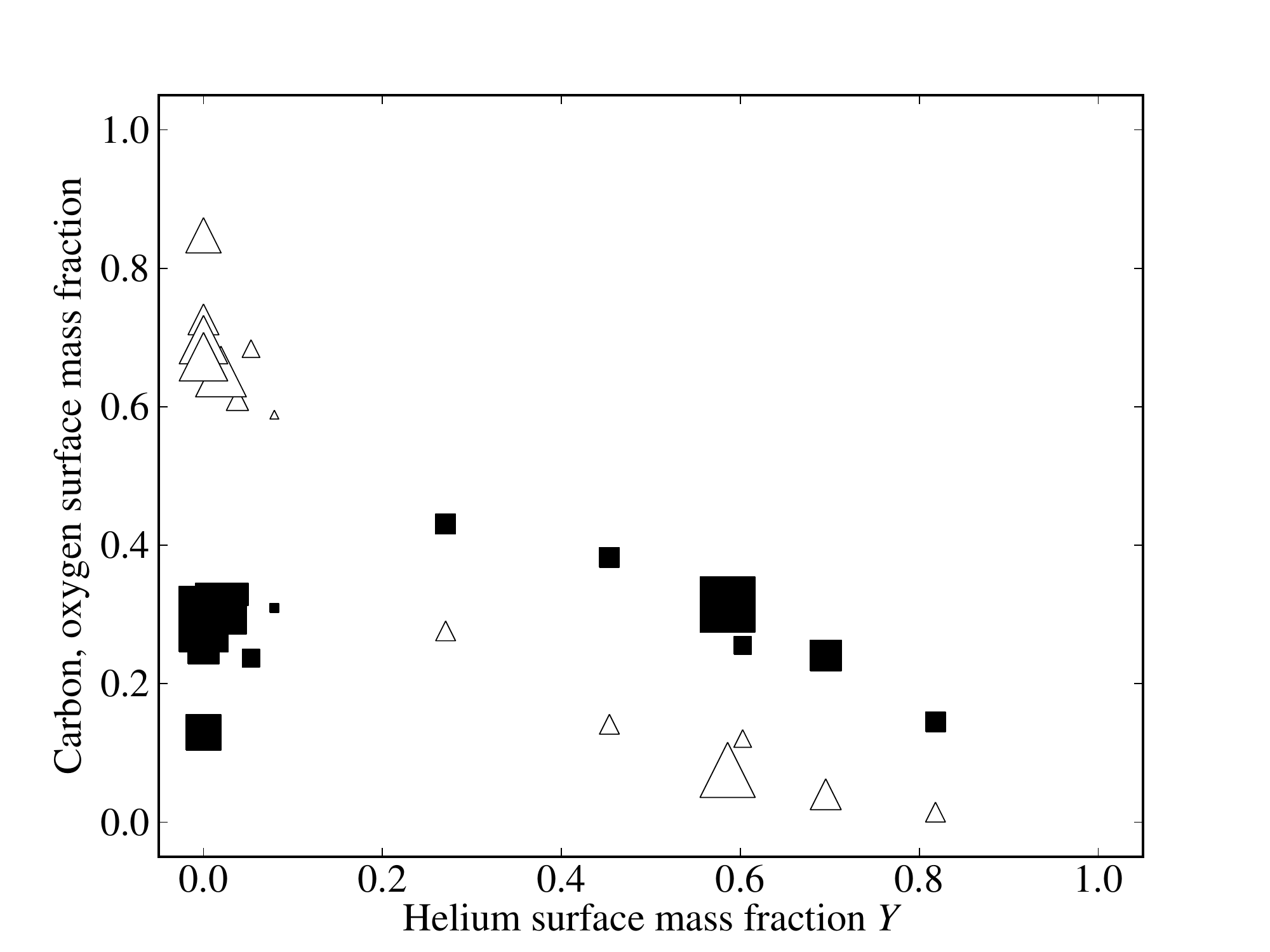}} 
\caption{Surface abundances of helium versus carbon (black squares) and oxygen (white triangles) for donors in the helium-star channel at the present time in the case of a sufficiently narrow star formation history. These correspond to the core abundances at the end of the tracks by \citet{nelemans2010}. The surface area of a symbol is proportional to the number of systems in the corresponding track.}
\label{fig:ucxb:surfcomp}
\end{figure}

The helium-star donors have partially turned into carbon-oxygen white dwarfs during their evolution, depending on their mass and evolutionary stage at the onset of Roche-lobe overflow (determined by the initial mass and orbital period). When the star starts mass transfer after filling its Roche lobe, burning is extinguished quickly \citep{savonije1986}, and at this stage the core mass fraction of helium varies between a few percent to almost $100\%$ \citep{nelemans2010}. Figure~\ref{fig:ucxb:surfcomp} shows the surface abundances at the present day, assuming a narrow star formation history ($\sigma = 0.5$ Gyr). Two thirds of the systems end up with less than $10\%$ helium on their surface. Systems that started out with a short orbital period generally have a higher helium mass fraction, because these had less time to burn helium before the onset of mass transfer. The abundances depend on the temperature at which helium and carbon burning takes place. A higher temperature causes a higher helium burning rate, producing more carbon. Later, the carbon abundance reduces in favor of oxygen. The scatter in Fig.~\ref{fig:ucxb:surfcomp} is therefore due to differences in core burning temperature caused by different stellar masses.

UCXBs produced via the white dwarf donor channel have donors mostly composed of either helium material (produced in the CNO cycle), or carbon-oxygen. The ratio between both types is about 3:1 but strongly depends on the efficiency of isotropic re-emission, which strongly affects the number of carbon-oxygen white dwarf donor systems that survive the onset of mass transfer (Sect.~\ref{sect:ucxb:wdire}).

Evolved main sequence donors with an initial mass of $1.0 - 1.2\ M_{\odot}$ reach a helium surface abundance $Y \approx 0.9$ once they become ultracompact, the remaining part being hydrogen and a metallicity $Z = 0.01$, which has been present from the start \citep{sluys2005a}.

Because the core material is exposed early on (e.g., at an orbital period shorter than $\sim \! 20$ min for the helium burning systems, \citealp{nelemans2010}), and the core is homogeneous due to convection during its burning stages, the chemical composition is expected not to change with increasing orbital period, and therefore the same for the total and the observable population.

\subsection{Collective emission as function of orbital period}

Even though variability behavior determines the number of UCXBs in outburst and their luminosity, the collective luminosity of all UCXBs at a given orbital period is in principle not dependent on variability, because the time-averaged mass transfer rate of an UCXB is a relatively straightforward function of orbital period.
During the evolution of an UCXB, its evolutionary timescale increases with age and orbital period. This means that there exist many more systems at longer orbital period. On the other hand the time-averaged mass transfer rate decreases with age and orbital period, so a long-period source has a lower time-averaged luminosity. The total energy output of a source, or of the population as a whole (under the assumption of a constant star formation rate), at a given orbital period is an indication of at which orbital periods systems are likely to be observed.

\begin{figure}
\resizebox{\hsize}{!}{\includegraphics{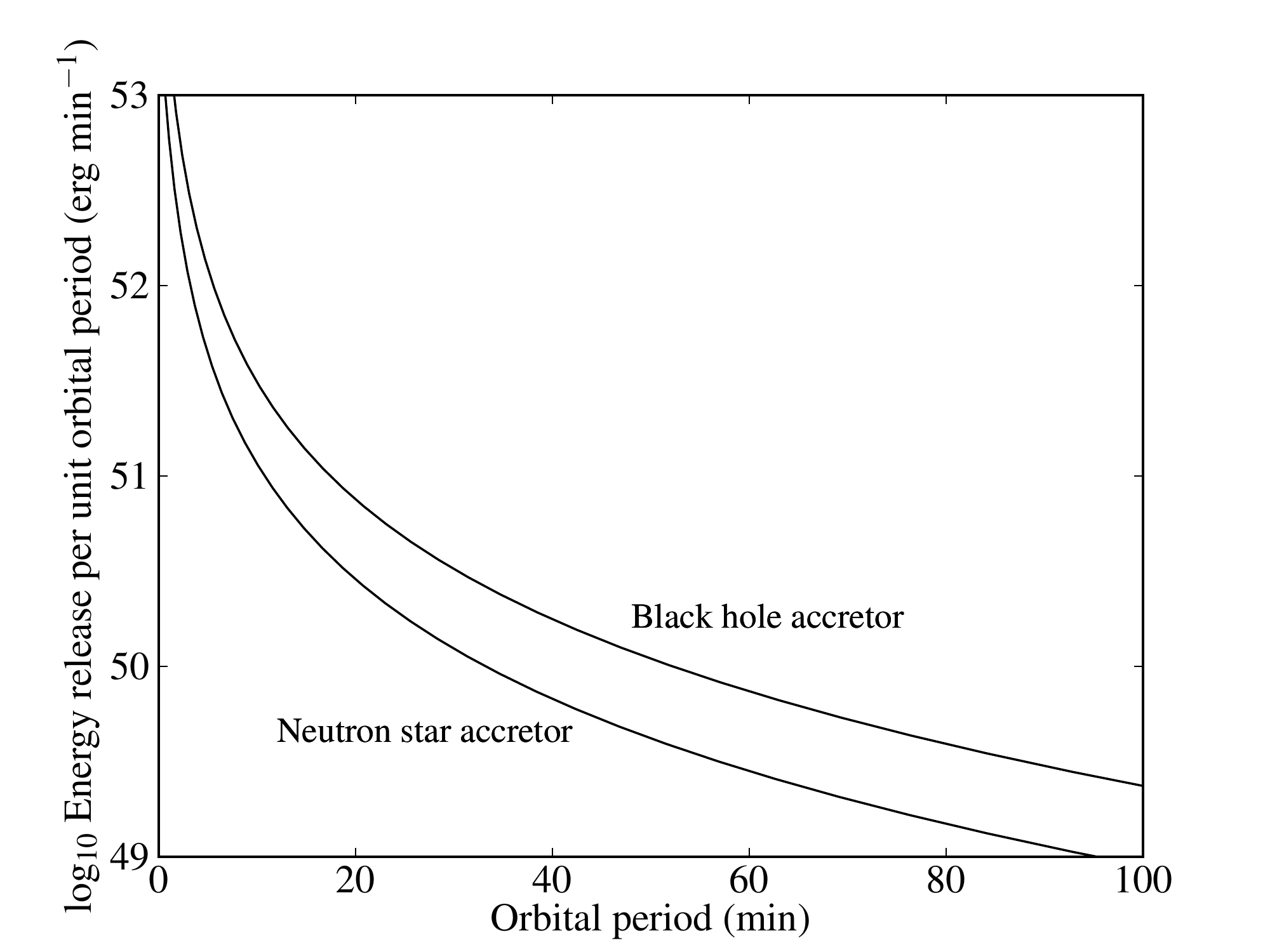}} 
\caption{Energy released per unit interval of orbital period for an UCXB with a zero-temperature helium white dwarf donor and a $1.4\ M_{\odot}$ neutron star accretor or a $10\ M_{\odot}$ black hole accretor. Note that the quantity on the vertical axis should not be interpreted as a luminosity; the time unit represents change in orbital period rather than passing time.}
\label{fig:ucxb:energyperiod}
\end{figure}

The amount of energy emitted by an UCXB per unit orbital period is given by
\begin{equation}
    \label{eq:ucxb:energyperiod}
    \frac{\mathrm{d} E}{\mathrm{d} P_\mathrm{orb}} = \frac{L}{\dot{P}_\mathrm{orb}} = -\frac{G M_\mathrm{a}}{2R_\mathrm{a}} \frac{\mathrm{d} M_\mathrm{d}}{\mathrm{d} P_\mathrm{orb}}
\end{equation}
where $E$ is the emitted energy and $M_\mathrm{d}$ the donor mass. This relation is illustrated in Fig.~\ref{fig:ucxb:energyperiod}. The donor mass decreases much faster at shorter orbital periods ($M_\mathrm{d} \propto P_\mathrm{orb}^{-1.3}$ \citep{vanhaaften2012evo}, i.e., $\mathrm{d} M_\mathrm{d} / \mathrm{d} P_\mathrm{orb} \propto P_\mathrm{orb}^{-2.3}$), and since the donor mass is the fuel for the luminosity, systems emit much more energy at short orbital periods, not only per time interval (their luminosity) but also per orbital period interval. For instance, an UCXB will emit $\sim \! 12$ times as much energy during its evolution from $20$ to $21$ min as it does between $60$ and $61$ min.
The consequence is that the short-period systems dominate the collective X-ray output of an UCXB population, unless the star formation rate decreases very fast.
Depending on the variability of systems and the sensitivity of the instrument used, this could very well result in short-period systems dominating the visible population.\footnote{Not to be confused with the known population, which is composed by individual observations at different times. In the known population, long-period systems will eventually dominate provided they can be seen during outbursts}

\section{Discussion}
\label{sect:ucxb:discussion}

We predict $\sim \! 1.9 \times 10^{5}$ of UCXBs in the Galactic Bulge, predominantly at orbital periods of $\gtrsim 70$ min, but also a few thousand systems with orbital periods shorter than $60$ min (but mostly longer than $40$ min).
Based on \textit{RXTE} ASM observations, about $40 - 80$ of these sub-hour UCXBs should be visible at high luminosities of $\gtrsim 10^{35}\ \mbox{erg s}^{-1}$ (Fig.~\ref{fig:ucxb:presentlum_asm}), depending on the star formation history. Also, $\sim \! 35 - 50$ bright UCXBs with orbital periods $\lesssim 30$ min (i.e., persistent) should be visible above such a luminosity based on gravitational energy release and the disk instability model (Figs.~\ref{fig:ucxb:presentlum_dim} and \ref{fig:ucxb:presentlum_dim2}).\footnote{The observed UCXBs with orbital periods around $20$ min are not clearly detected by ASM most of the time, even though they are predicted to be persistent by the disk instability model. Incorporating the fact that they are bright less than $100\%$ of the time, the predicted number of bright systems becomes lower, depending on the precise luminosity distribution of these systems.}

The combined common-envelope parameter $\alpha_\mathrm{CE} \lambda$ for massive stars may be lower than the value of $2$ that we used \citep{voss2003}. Decreasing this value to $0.2$, as well as using a Maxwellian kick velocity distribution with a dispersion of $450\ \mbox{km s}^{-1}$, rather than the distribution by \citet{paczynski1990}, would reduce the number of UCXBs formed by a factor of $\sim \! 8$, as fewer systems will survive the common-envelope stage or the supernova explosion. In that case, the total number of UCXBs in the Bulge we predict is $\sim \! 2 \times 10^{4}$, and the number of bright systems becomes $\sim \! 5 - 10$. Table \ref{table:ucxb:params} shows the number of UCXBs in our model for various combinations of common-envelope efficiency and neutron star kick velocity distribution.
Furthermore, the slope of the initial mass function at high stellar mass is also uncertain. A steeper slope (resulting from earlier studies such as \citealt{kroupa1993}) leads to a smaller fraction of massive stars and therefore fewer UCXBs. A different choice for the initial component mass pairing may also reduce the number of UCXBs by an order of magnitude \citep{belczynski2004popsynt}.

\begin{table}
\centering
\caption{Size of the modeled UCXB population in the Galactic Bulge for different model parameters.}
\label{table:ucxb:params}
\begin{tabular}{ccc}
\hline\hline
$\alpha_\mathrm{CE} \lambda$      & Kick distribution      & Number of UCXBs ($\times\ 10^{4}$)       \\
\hline
$2$            & Paczy\'{n}ski       & 19        \\ 
$2$            & Maxwellian          & 14        \\ 
$0.2$          & Paczy\'{n}ski       & 4         \\ 
$0.2$          & Maxwellian          & 2         \\ 
\hline
\end{tabular}
\end{table}

We can distinguish several disagreements with observations. First, no UCXBs with orbital periods longer than $60$ min have been discovered, faint or bright, in the Bulge or elsewhere. Second, no bright UCXBs with a short orbital period ($\lesssim 30$ min) have been identified in the Galactic Bulge. Third, only three UCXBs with orbital periods between $40$ and $55$ min, \object{XTE J1807--294} \citep{markwardt2003b}, \object{XTE J1751--305} \citep{markwardt2002} and \object{SWIFT J1756.9--2508} \citep{krimm2007}, are presumably located in the Bulge, based on their positions in the sky, as their distances are not known.

As for the predicted $\sim \! 1.9 \times 10^{5}$ long-period systems, the probable existence of three observed UCXBs with orbital periods shorter than $55$ min in the Bulge can be used to calibrate the formation rate of UCXBs, \textit{independent of population synthesis}. This yields a much larger number of UCXBs than three for systems with a period longer than $55$ min, based only on the rapid increase of the evolutionary timescale (set by gravitational wave radiation) with orbital period. For instance, UCXBs are expected to reach an orbital period of $55$ min within $\sim \!2$ Gyr after the onset of mass transfer \citep{vanhaaften2012evo}.

Even though the disk instability model predicts these long-period systems to be in outburst so rarely that few are expected to be bright, in quiescence they would still be detectable by sensitive instruments at $\sim \! 10^{31-33}\ \mbox{erg s}^{-1}$ \citep[e.g.][]{bildsten2001proceed,heinke2003,belczynski2004center}.

The three observed UCXBs that are located in the direction of the Bulge have undetermined donor compositions, though \object{SWIFT J1756.9--2508} is thought to have helium composition \citep{krimm2007}. This cannot be used to constrain the star formation history, as UCXBs with helium white dwarf donors can form with a wide range of delay times.

\subsection{Comparison with previous studies}

\citet{belczynski2004popsynt} performed a population synthesis study of (primordial) UCXBs in the Galactic disk. The main difference between their and our results is that they found a total of $478$ UCXBs with orbital periods shorter than $80$ min in the disk at the present epoch, about three orders of magnitude fewer than our result, per unit star forming mass. It is not clear what causes this discrepancy, although these authors use different initial binary parameters than we do (Appendix \ref{sect:ucxb:appimf}), for example a steeper high-mass slope in the initial mass function that leads to fewer massive stars, relatively. Also, their initial primary masses leading to UCXBs span a narrower range.
Of all UCXBs in their simulation, about $20\%$ have a black hole accretor, all of which form via the accretion-induced collapse of a neutron star. This percentage strongly depends on the assumed upper mass of a neutron star ($2\ M_{\odot}$), the mass retention efficiency of the accreting neutron star and the evolutionary stage at which the common envelope happens. As in our study, these authors did not find any (surviving) UCXBs with a black hole that was formed directly in the collapse of a massive star. They found that about $90\%$ of the neutron star accretors form via the accretion-induced collapse of an oxygen-neon-magnesium white dwarf, a scenario our model does not produce.
In our simulations, $81\%$ of the UCXBs start mass transfer from a helium burning donor, compared to $40\%$ in \citet{belczynski2004popsynt} for the Galactic field. This is not unreasonable given the uncertainties in e.g. the onset of mass transfer from a white dwarf, and differences in assumptions between both studies.
The number of persistent sources predicted by the disk instability model depends sensitively on the orbital period separating the persistent and transient sources, because most of the predicted persistent sources have orbital periods only slightly shorter than this critical period. Details in accretion disk models (e.g. X-ray irradiation) and composition can make a large difference. We predict about $0.02\%$ of the UCXBs to be persistent (Sect.~\ref{sect:ucxb:resdim}), a much smaller fraction than found by \citet{belczynski2004popsynt} ($2.2\%$), but their number applies to the Galactic disk, which has a several orders of magnitude higher ongoing star formation rate, and therefore more young UCXBs, which have short orbital periods.

Our number of persistently bright UCXBs ($35 - 50$ from the disk instability model, using our standard parameters) can also be compared with the number of persistent UCXBs with white dwarf donors ($600 - 900$) predicted by \citet{zhu2012} for the whole Galaxy, if one takes into account the difference in adopted cut-off donor mass for persistent behavior. These authors found that UCXBs with donor masses lower than $0.03\ M_{\odot}$ are transient whereas our limit lies around $0.02\ M_{\odot}$. Using their limit, our estimate would reduce to roughly ten, which scales within a factor of a few with their number given the stellar mass ratio between Bulge ($1 \times 10^{10}\ M_{\odot}$) and Galaxy (an additional $4 - 6\ \times 10^{10}\ M_{\odot}$ in the Disk, \citealp{klypin2002}).\footnote{\citet{zhu2012} found a much larger number of UCXBs with a helium burning donor than our study, but this is expected given the large difference in recent star formation between Bulge and Disk.} The overprediction of bright, persistent UCXBs is therefore not unique to our study.

Recently, \citet{zhu2012ucxb} performed a population synthesis study of Galactic UCXBs with neutron star accretors, and predicted $5 - 10\ \times 10^{3}$ systems in the Galaxy, depending on neutron star birth kicks. As in our study, the helium burning donor channel was the most common. Notable differences with our work are that these authors found a large number of UCXBs with a carbon-oxygen white dwarf origin, and a peak in the orbital period distribution near 40 min.

\subsection{Overprediction of UCXBs with long orbital period}
\label{sect:ucxb:disclong}

An important clue towards what may happen at long periods comes from the long-term ASM data. The reason for the difference in predictions by the ASM and disk instability model (Sect.~\ref{sect:ucxb:resobs}) lies in the ASM observations that the UCXBs with orbital periods $40 - 55$ min are approximately two orders of magnitude brighter than theoretically expected from the time-averaged mass transfer rate, assuming mass transfer is driven exclusively by gravitational wave radiation in a binary with a (semi-)degenerate donor \citep{vanhaaften2012xlum}.\footnote{These ASM luminosities are direct observations, unrelated to the extrapolations of the curves shown in Fig.~\ref{fig:ucxb:signilum} and discussed in Sect.~\ref{sect:ucxb:resobs}.}
Assuming that the observed systems have been displaying normal behavior during the $16$ years of \textit{RXTE} observations, additional angular momentum loss besides that from gravitational wave radiation would cause a higher mass transfer rate at the same orbital period, and therefore a higher time-averaged luminosity \citep[see also][]{ruderman1989late}.
As mentioned in Sect.~\ref{sect:ucxb:old}, an efficient physical mechanism for additional loss of angular momentum from the system is a wind from the donor, induced by irradiation from the accretion disk or millisecond pulsar. In black widow systems, which host a millisecond pulsar and a low-mass ($\lesssim 0.2\ M_{\odot}$) companion in a $< 10$ h orbit \citep{king2005}, such donor evaporation has been observed \citep[e.g.][]{fruchter1988}. This scenario has also been proposed to be happening to the unusually light ($\sim \! 10^{-3}\ M_{\odot}$) detached companion to the millisecond pulsar \object{PSR J1719--1438} \citep{bailes2011} via either the white dwarf or helium burning donor channels \citep{vanhaaften2012j1719} or the evolved main sequence donor channel (with an orbital period minimum at $\sim \! 45$ min, \citealp{benvenuto2012}), though the latter scenario does not produce a carbon-oxygen rich donor.

The recently discovered spin-powered millisecond gamma-ray pulsar \object{PSR J1311--3430} system \citep{pletsch2012}, with an orbital period of $93.8$ min \citep{romani2012,kataoka2012} and an evaporating helium donor \citep{romanietal2012} supports the hypothesis that UCXB evolution is strongly influenced by donor evaporation. Given the low hydrogen abundance, donor evaporation, orbital period, pulsar spin period, and minimum companion mass, PSR J1311--3430 could very well be an UCXB descendant on its way to becoming a millisecond radio pulsar system like PSR J1719--1438.

The non-detection of UCXBs with periods longer than $\sim \! 60$ min, when the donor is still expected to be much more massive than the companion in \object{PSR J1719--1438}, suggests the existence of another mechanism that hides UCXBs with low donor masses and low mass transfer rates (the $60$ min limit is uncertain due to the small observed sample). The propeller effect (Sect.~\ref{sect:ucxb:old}) is the most promising mechanism to explain this non-detection. The rotational energy of the millisecond pulsar is sufficient to make long-period UCXBs with very low mass transfer rates ($\sim \! 10^{-13}\ M_{\odot}\ \mathrm{yr}^{-1}$) much fainter, since it causes arriving matter to be unbound \citep{vanhaaften2012evo}. The propeller effect could still allow for a (very) low rate of accretion that would prevent radio emission and make the sources visible in the ultraviolet (owing to their low disk temperatures and possibly disturbed inner accretion disks).
Furthermore, radio emission from a millisecond pulsar itself, once switched on after an interruption in mass transfer, is capable of preventing accretion \citep{burderi2001,fu2011}.

Using the \textit{Chandra X-Ray Observatory} \citep{weisskopf2002}, the Galactic Bulge Survey has found $1234$ X-ray sources in $8.3\ \mbox{deg}^{2}$ \citep{jonker2011} so far, most of which have not yet been identified. Although many are expected to be foreground Cataclysmic Variables or non-ultracompact X-ray binaries, this number of systems found in approximately $5\%$ of the total area of the Bulge on the sky is at least consistent with a large population of faint X-ray binaries.
Also, a potentially large population of sub-luminous X-ray transients with neutron star accretors exists near the Galactic Center \citep{sakano2005,wijnands2006,degenaar2009,degenaar2010}. These systems have (intrinsic) peak luminosities near $\sim \! 10^{34-35}\ \mbox{erg s}^{-1}$ (in the $2 - 10$ keV range), and may include UCXBs, although the disk instability model predicts peak luminosities $\gtrsim \! 10^{37}\ \mbox{erg s}^{-1}$ for UCXBs.
\citet{king2006} found that the luminosities of some very faint X-ray transients imply mass transfer rates of $\sim \! 10^{-13}\ M_{\odot}\ \mathrm{yr}^{-1}$, which is consistent with the behavior of old UCXBs.

The additional angular momentum loss increases the time derivative of the orbital period, and as a result the actual number of systems at long periods in our prediction based on only gravitational wave radiation (Figs.~\ref{fig:ucxb:presentpop} and \ref{fig:ucxb:presentpop2}) should be reduced by two orders of magnitude.

\subsection{Overprediction of UCXBs with short orbital period}

Since UCXBs with an orbital period shorter than $\sim \! 30$ min are expected to be persistently bright, our overprediction of these binaries ($\sim \! 5 - 50$ systems based on the disk instability model, depending on assumptions in the model, is about one order of magnitude more than the three observed Bulge UCXBs) can have several causes: the population synthesis model produces too many UCXBs, fewer UCXBs survive the onset of mass transfer, or short-period UCXBs are bright less than $100\%$ of the time.

It is uncertain whether the white dwarf donor mass limit of $0.38\ M_{\odot}$ for isotropic re-emission (based on the zero-temperature mass-radius relation and a consideration of the energy necessary to eject matter) should be used as the threshold for the survival of a system. A different assumption in the details of the hydrodynamics of the onset of mass transfer could result in either a lower or a higher limit. However, the number of surviving white dwarf donors is not very sensitive to small changes in the isotropic re-emission limit because of the small number of donors around this value (Fig.~\ref{fig:ucxb:hist_wd}). On the other hand, due to the sensitivity of this donor mass limit to the actual and critical mass transfer rates \citep{vanhaaften2012evo}, the number of surviving UCXBs could be significantly smaller if the isotropic re-emission efficiently is less than unity, which seems plausible.
However, systems could survive a mass transfer rate exceeding the isotropic re-emission limit if an UCXB with a $\sim \! 2$ min orbital period is able to survive a few hundred years ($\sim \! 10^{8}$ orbits) with a significant amount of matter orbiting in and around the binary, perhaps in a circumbinary ring \citep{soberman1997,ma2009} that would be removed at a later stage.
The value of isotropic re-emission limit can be tested only using short-period UCXBs, because for those the white dwarf donor channel is expected to dominate. In the total population (mostly long-period systems), the helium donor channel is more important. This channel does not experience the high mass transfer rates that characterize the onset of mass transfer from a white dwarf donor. Near the period minimum the mass transfer rate remains below approximately three times the Eddington limit (Fig.~\ref{fig:ucxb:hetracks})

Even though Fig.~\ref{fig:ucxb:history} shows that the contribution of the helium-star channel dominates, its importance has not been established observationally yet, as no detached (short orbital period) helium star--neutron star binaries have been discovered so far \citep{nelemans2010}. If helium burning stars would turn into white dwarfs before the onset of Roche-lobe overflow, over $90\%$ would be unable to survive as a binary system.

Given these uncertainties, our overprediction by approximately one order of magnitude may simply be a consequence of poorly known parameters in our simulation, and in this case \textit{no problematic discrepancy} between our results and X-ray observations would remain.

\subsection{Consequences for the population of millisecond radio pulsars}

If mass transfer would cease completely in most old UCXBs as suggested by the existence of \object{PSR J1719--1438}, then a fraction of the neutron stars they harbor would become visible to us as (binary) millisecond radio pulsars for several billion years. The same is probably true for UCXBs in environments closer to us than the Bulge. The number of isolated or binary millisecond radio pulsars in the Bulge suggested by the number of UCXBs with $\gtrsim 70$ min periods we predict in our standard model, after correcting for the factor of ten overestimation in the UCXB birth rate identified above, is about $2 \times 10^{4}$, assuming the pulsars have not yet turned off as a result of spinning down. The estimated \textit{Galactic} population of millisecond pulsars, based on the observed population, is $4 \times 10^{4}$ \citep{lorimer2008}. Of the known Galactic Disk population (i.e., excluding globular clusters) of $\sim \! 100$ radio pulsars with spin periods shorter than $10$ ms, about $60\%$ have a companion too massive to be consistent with late-time UCXB evolution, based on the \textit{ATNF Pulsar Catalogue}\footnote{\url{http://www.atnf.csiro.au/people/pulsar/psrcat/}} in January 2013 \citep{manchester2005}. (These have descended from hydrogen-rich low-mass and intermediate-mass X-ray binaries, \citealp{bhattacharya1991,tauris2011}.) Extrapolating this to the estimated Galactic population, $\sim \! 2 \times 10^{4}$ millisecond pulsars are left that have no or a very low-mass companion, roughly the same number we predict from UCXBs for the Bulge alone (after normalizing the number of short-period UCXBs).

It seems likely that millisecond pulsars have evaporated their companion entirely and are left as isolated millisecond pulsars given the reasonable match between the predicted number of old UCXBs and the number of isolated millisecond radio pulsars, combined with the very small number of observed millisecond radio pulsars with companions with masses lower than $0.01\ M_{\odot}$.
Alternative formation channels for isolated millisecond pulsars are spin up of a neutron star by a disrupted white dwarf companion \citep{vandenheuvel1984}, and disruption of a millisecond pulsar binary during the supernova explosion of the donor star in a high-mass X-ray binary \citep[e.g.][]{camilo1993}. The number of neutron star--white dwarf mergers, however, seems too high to be consistent with the number of isolated millisecond pulsars that have formed. Based on Fig.~\ref{fig:ucxb:hist_wd}, merging systems are much more common than surviving UCXBs, also after including UCXBs from the helium burning donor channel \citep[see also][]{iben1995}. On the other hand, the number of millisecond pulsars that lose their companions when it explodes as a supernova seems too small to be responsible for a large fraction of the isolated millisecond pulsars \citep{burgay2003,belczynski2010} -- moreover the high-mass donor star may not live long enough to spin up the neutron star to a spin period shorter than $\sim \! 10$ ms.

\section{Summary and conclusions}
\label{sect:ucxb:conclusion}

We modeled the present-day population of primordial ultracompact X-ray binaries in the Galactic Bulge with the purpose of gaining insight in their formation and evolution. Both binary evolution and accretion physics determine the observable population, and we attempted to disentangle these in this study. We considered three main formation channels: systems that start Roche-lobe overflow by a white dwarf donor, a helium burning donor or an evolved main sequence donor. Our simulations have not produced UCXBs containing a black hole, because most systems with a very massive primary merge during unstable mass transfer, and the small number that remains is expected to merge during the onset of mass transfer from a relatively massive white dwarf to the black hole. Thus, all UCXB systems in our simulations have a neutron star accretor.

The vast majority of UCXBs form via the helium burning donor channel ($81\%$) or the white dwarf donor channel ($18\%$), and therefore their exposed cores are expected to show either carbon and oxygen in their spectra, or helium, as well as small amounts of other reaction products. These two channels differ in the delay time between the zero-age main sequence and the onset of Roche-lobe overflow to the neutron star. In the white dwarf channel this delay can be as long as the age of the Universe or more (though for most systems it is less than a few billion years), whereas in the helium burning channel the delay is less than $1$ Gyr.

The size and characteristics of the present-day population are only marginally dependent on the assumed width, $\sigma$, of the Gaussian distribution describing the star formation history if this value is $\lesssim 1$ Gyr. This is because these values of $\sigma$ are small compared to the age of a $10$ Gyr old system. A broad star formation history allows for recent star formation and short orbital period UCXBs with a helium burning donor origin, because of their short delay time. With a narrow star formation history, short-period UCXBs must have a white dwarf origin and therefore can have helium composition.

Very short period UCXBs can have a helium or carbon-oxygen white dwarf donor, since these must have formed recently. Recent UCXB formation is dominated by the white dwarf donor channel, even for $\sigma = 2.5$ Gyr.

The number of predicted systems with orbital periods shorter than $\sim \! 30$ min is particularly important, since those systems are probably observable as persistently bright sources, and therefore well suited to test and calibrate the simulations. We predict about $40$ bright sources, mostly of helium and carbon-oxygen composition and with orbital periods shorter than $30$ min. The UCXBs with the shortest periods ($\lesssim 20$ min) are more likely to have helium composition. The observed number of bright UCXBs is about ten times smaller than suggested by our model, which reflects the uncertainties in the adopted star formation history, initial binary parameters, natal kick velocities of neutron stars, common-envelope parameters and the onset of mass transfer to a neutron star accretor.

We predict about $(0.2 - 1.9) \times 10^{5}$ UCXBs in the Galactic Bulge, and we stress that such a large population is necessary based on the simple argument that the evolutionary timescale of UCXBs increases rapidly towards longer orbital periods, and therefore the observed number of short-period UCXBs, in the Bulge and also in the Galactic Disk, implies several orders of magnitude more UCXBs at long orbital periods ($> 60$ min). With different model assumptions, this number could be up to an order of magnitude lower.

Irradiation of the donor star by the neutron star and accretion disk strongly influences UCXB evolution, at least at orbital periods longer than $40$ min. These systems \textit{evolve much faster}, probably by $\sim \! 100$ times, than they would if their evolution was driven exclusively by angular momentum loss via gravitational wave radiation, as assumed in this paper. UCXBs with orbital periods longer than $1$ h have not been detected yet, which implies that, if existent, these systems are very faint in all electromagnetic bands (and therefore cannot be considered true X-ray binaries). We suggest that the majority of these systems have orbital periods on the order of $1.5 - 2.5$ h rather than the $\sim \! 1.3$ h expected from gravitational wave driven evolution. Furthermore we expect that the neutron stars have companions with masses much lower than $0.01\ M_{\odot}$, and could very well have evaporated their companions entirely, being left as isolated millisecond pulsars.

In a forthcoming paper we will model the population of hydrogen-rich low-mass X-ray binaries in the Galactic Bulge.

\begin{acknowledgements}
LMvH, GN, RV, and SFPZ are supported by the Netherlands Organisation for Scientific Research (NWO). GN and RV are supported by NWO Vidi grant $\#016.093.305$ to GN. SFPZ is supported by NWO grants $\#639.073.803$ (Vici) and $\#614.061.608$, and the Netherlands Research School for Astronomy (NOVA). LRY is supported by RFBR grant \#10-02-00231 and the Program P-21 of the Praesidium of Russian Academy of Sciences. This research has made use of NASA's Astrophysics Data System Bibliographic Services (ADS).
\end{acknowledgements}

\appendix

\section{Binary initial mass function and normalization of the simulation}
\label{sect:ucxb:appimf}

In the initial binary system, the more massive component is called the primary. We use primary-constrained pairing to construct `zero-age' binaries \citep{kouwenhoven2008}. The primary masses $M_\mathrm{primary}$ of the zero-age main sequence binaries are drawn from the stellar initial mass function (IMF) of \textit{primaries} in massive star clusters that we derive from the results by \citet{kroupa2001}, where $M$ is the stellar mass and $0.08 \leq M/M_{\odot} \leq 100$. The mass ratio $0 < M_\mathrm{secondary}/M_\mathrm{primary} \le 1$ of the components is subsequently drawn from a constant distribution \citep{kraicheva1989,hogeveen1992} -- secondary masses lower than $0.08\ M_{\odot}$ are accepted.
The eccentricity $e$ distribution is proportional to $e$ between $0$ and $1$, and the semi-major axis $a$ distribution is inversely proportional to $a$ \citep{popova1982,abt1983}, up to $10^{6}\ R_{\odot}$ \citep{duquennoy1991} -- the lower limit is set by the requirement that the initial stellar radii fit inside the circularized orbit.

The specific binary fraction as a function of $M$ is given by the observationally practical definition \citep{reipurth1993,kouwenhoven2009}
\begin{align}
    \label{eq:ucxb:binaryfraction}
    \begin{split}
    \mathcal{B}(M) \equiv\ & \frac{N_\mathrm{binary}(M_\mathrm{primary} = M)}{N_\mathrm{single}(M) + N_\mathrm{binary}(M_\mathrm{primary} = M)} \\
    =\ & \frac{N_\mathrm{binary}(M_\mathrm{primary} = M)}{\mathit{IMF}(M)}
    \end{split}
\end{align}
where $N_\mathrm{single}(M)$ is the distribution of single stars of mass $M$, $N_\mathrm{binary}(M_\mathrm{primary} = M)$ the distribution of binary systems containing a primary of mass $M$, and $\mathit{IMF}(M)$ the IMF of \textit{systems} (single stars and multiple systems combined) by \citet{kroupa2001}. Based on observations summarized in \citet{kouwenhoven2009,kraus2009,sana2012} we approximate
\begin{equation}
    \label{eq:ucxb:binarity}
    \mathcal{B}(M) = \frac{1}{2} + \frac{1}{4}\log_{10}(M) \qquad  (0.08 \le M/M_{\odot} \le 100)
\end{equation}
where we assume all multiple systems to be binaries.
Equation~(\ref{eq:ucxb:binaryfraction}) can be separated as
\begin{align}
    \label{eq:ucxb:nsin}
    \begin{split}
    N_\mathrm{binary}(M_\mathrm{primary} = M) \propto\ & \mathcal{B}(M)\ \mathit{IMF}(M), \\
    N_\mathrm{single}(M) \propto\ & [1 - \mathcal{B}(M)]\ \mathit{IMF}(M).
    \end{split}
\end{align}
It follows that single stars are more common than binary systems; there are $1.6$ single stars for each binary system. The mass per binary system including the corresponding single stars (which can be a fractional number) is given by
\begin{equation}
    \label{eq:ucxb:totmass}
    M_\mathrm{T} = \frac{1}{\mathcal{B}(M_\mathrm{primary})} M_\mathrm{primary} + M_\mathrm{secondary},
\end{equation}
and the average star forming mass for each binary system formed (i.e., including mass from single stars) by
\begin{equation}
    \label{eq:ucxb:avgtotmass}
    \bar{M}_\mathrm{T} = \int_{0.08\ M_{\odot}}^{100\ M_{\odot}} \left( 1 + \frac{\mathcal{B}(M)}{2} \right) \mathit{IMF}(M) M \mathrm{d}M \approx 1.9\ M_{\odot}
\end{equation}
(the factor $1/2$ appears because the average secondary mass is equal to half of the average primary mass for the chosen constant mass ratio distribution). This number is the sum of the average primary mass ($0.86\ M_{\odot}$),\footnote{This value is higher than the average mass of $0.57\ M_{\odot}$ of the IMF by \citet{kroupa2001} because single stars are excluded.} the average secondary mass ($0.43\ M_{\odot}$) and the corresponding average mass in single stars per binary system ($0.64\ M_{\odot}$). A lower limit of $0.1\ M_{\odot}$ increases the average mass per binary by $\sim \! 12\%$. Overall two-thirds of the star-forming mass is in binaries. The total number of binaries that forms in the Galactic Bulge is normalized using the total number of stars
\begin{equation}
    \label{eq:ucxb:nbin}
    \int_{0.08\ M_{\odot}}^{100\ M_{\odot}} N_\mathrm{binary}(M)\, \mathrm{d}M = \frac{1 \times 10^{10}\ M_{\odot}}{\bar{M}_\mathrm{T}} \approx 5.2 \times 10^{9}.
\end{equation}
Of all primaries, $1.3\%$ have a mass higher than $8\ M_{\odot}$. For these masses, the power-law slope of the primary IMF (defined over linear mass intervals), from which we draw primary masses, varies between $-2.15$ (for $M = 8\ M_{\odot}$) and $-2.2$ ($M = 100\ M_{\odot}$), compared to the estimate of $-2.3$ by \citet{kroupa2001} for the combined IMF of single stars and primary components.
The IMF of primary components $N_\mathrm{binary}(M_\mathrm{primary} = M)$ is flatter than the IMF of systems $\mathit{IMF}(M)$ because Eq.~(\ref{eq:ucxb:binarity}) is an increasing function (most low-mass stars are single whereas massive stars are usually in binaries).\footnote{A non-zero mass-\textit{independent} binary fraction leads to an IMF of all stars combined that is steeper than the IMF of systems \citep{sagar1991,scalo1998,kroupa2001}. This does not affect our method as we only consider the IMF of all primary components.}
The IMF for single stars only is steeper than $-2.3$ and steepens towards high mass.

\bibliographystyle{aa}
\bibliography{lennart_refs}

\end{document}